\documentclass[floatfix,prd,twocolumn,letterpaper,lengthcheck,superscriptaddress,showpacs,amssymb,amsmath,amsfonts,aps,altaffilletter,nofootinbib,nopreprintnumbers,longbibliography]{revtex4-1}

\usepackage{color}
\usepackage{hyperref}
\usepackage{amsmath}
\usepackage{amsthm}
\usepackage{multirow}
\usepackage{graphicx}
\usepackage{tikz}
\usetikzlibrary{calc}
\usepackage{placeins}
\usepackage{xspace}
\usepackage{physics,enumerate}

\usepackage{mathrsfs}
\DeclareSymbolFontAlphabet{\mathrsfs}{rsfs}

\newcommand{\MM}{\ensuremath{\mathcal{M}}} 

\newcommand{\Surf}{\ensuremath{\mathcal{S}}}

\newcommand{\Kpp}{R^2 k_{\phi \phi}}

\newcommand{\tq}{{q}}

\def\MADM{M_\text{ADM}}

\newcommand{\Lslice}{L_\Sigma} 

\newcommand{\eqz}{\overset{z}{=}}
\newcommand{\eqg}{\overset{\gamma}{=}}

\def\MOTSs{MOTSs\xspace}   
\def\MOTOSs{MOTOSs\xspace} 

\begin{document}

\title[]{Ultimate fate of apparent horizons during a binary black hole merger I:\\
locating and understanding axisymmetric marginally outer trapped surfaces}

\author{Ivan Booth}
\affiliation{
    Department of Mathematics and Statistics, Memorial University of Newfoundland,
    St. John's, Newfoundland and Labrador, A1C 5S7, Canada
}

\author{Robie A. Hennigar}
\affiliation{
    Department of Mathematics and Statistics, Memorial University of Newfoundland,
    St. John's, Newfoundland and Labrador, A1C 5S7, Canada
}
\affiliation{
	Department of Physics and Astronomy, University of Waterloo, 
	Waterloo, Ontario, Canada, N2L 3G1
}
\affiliation{
	Department of Physics and Computer Science, Wilfrid Laurier University, 
	Waterloo, Ontario, Canada N2L 3C5
}	

\author{Daniel Pook-Kolb}
\affiliation{
    Max-Planck-Institut f\"ur Gravitationsphysik (Albert Einstein Institute),
    Callinstr. 38, 30167 Hannover, Germany
}
\affiliation{
    Leibniz Universit\"at Hannover, 30167 Hannover, Germany
}


\begin{abstract}

    In classical numerical relativity, marginally outer trapped surfaces
    (\MOTSs) are the main tool to locate and characterize black holes.
    For five decades it has been known that during a binary merger, 
    a new outer horizon forms around the initial apparent horizons of the 
    individual holes once they are sufficiently close together. 
    However the ultimate fate of those initial horizons has remained a subject
    of speculation. Recent axisymmetric studies have shed new light on this 
    process and this pair of papers essentially completes that line of research: we resolve
    the key features of 
    the post-swallowing axisymmetric evolution of the
    initial horizons. This first paper introduces a new shooting-method for 
    finding axisymmetric \MOTSs along with a reinterpretation of the stability 
    operator as the analogue of the Jacobi equation for families of \MOTSs. 
    Here, these tools are used to study exact solutions and initial data. In the 
    sequel  paper \cite{PaperII} they are applied to black hole mergers.      
    
%
%
%
    
\end{abstract}

\maketitle

\section{Introduction}
\label{sec:intro}

Since the early 1970s it has been known that during a binary black hole collision 
the event horizons of the original black holes merge via the ``pair of pants'' diagram\cite{hawking_ellis_1973}. 
As the evolution progresses, the initially distinct event horizons approach and then touch at a caustic point. Subsequently,
that point opens up and a new, merged, event horizon remains. 

The picture for apparent horizons is more complicated, though
the initial stages of this evolution have also been 
understood for many decades\cite{hawking_ellis_1973}. When the black holes become sufficiently close,
a new apparent horizon instantaneously forms around and outside of the original pair (this is often referred to as an 
apparent horizon jump). This immediately splits into an outer apparent horizon and an inner horizon-like structure  
which respectively move outwards 
and inwards. The original horizons remain inside. The outer apparent horizon and event horizon asymptote
towards each other but the ultimate fate of the original apparent horizons and inner horizon-like structure has remained unresolved.

Before continuing, note that the term ``apparent horizon'' has accreted several distinct, though closely related, usages 
over the last five decades. It is sometimes used as a synonym for \emph{marginally outer trapped surface} (MOTS): 
a closed surface of vanishing outward null expansion\footnote{
The original definition of an apparent horizon (see, for example, \cite{hawking_ellis_1973}) was as the boundary of the 
trapped region in a time slice $\Sigma_t$. Then it was proved that such a boundary is necessarily a MOTS. However it 
is certainly not true that every \MOTSs is the boundary of some trapped region. 
}. However we will reserve it for \MOTSs that can reasonably be thought of as black hole boundaries: stable \MOTSs
in the sense of \cite{Andersson:2005gq,Andersson:2007fh} (discussed in more detail in Section \ref{sub:stabilityTheory} of this paper) which intuitively can be thought of as  \MOTSs that are (or were at some point in the past) outermost in 
the foliation in which we are working. 
Hence in the case 
of a binary merger we would continue to refer to both the original and outermost \MOTSs as 
apparent horizons but the inner horizon-like structure will now just be a MOTS. 

Any three-surface foliated by \MOTSs will be referred to as a \emph{marginally outer trapped tube} (MOTT) but 
if those \MOTSs are apparent horizons we will refer to it as a \emph{dynamical apparent horizon} (DAH)\footnote{
This name is inspired by, though distinct from, the dynamical horizons of \cite{Ashtekar:2003hk} which were 
spacelike MOTTs with strictly negative inward null expansions. 
More recently dynamical horizon has also been used as a synonym for MOTTs\cite{pook-kolb2020I,pook-kolb2020II}. 
These definitions do not refer to a spacetime foliation. By contrast our DAHs are foliation dependent. However the definition
has been  adapted 
to match the much wider range of MOTS now known to exist. 
}. The philosophy behind this naming is to reserve the term 
``horizon'' for objects that can reasonably be thought of as black hole boundaries. The reason for this restriction will 
soon become clear: for every spacetime that we have studied we have found an apparently infinite number 
of \MOTSs/MOTTs. As we shall see, while they are crucial for understanding black hole mergers, these exotic \MOTSs 
are not in any sense black hole boundaries.

%

We return to mergers. Following the appearance of the inner and outer \MOTSs, 
it is now understood that the apparent horizons of the original black holes 
interpenetrate \cite{Szilagyi_2007,Moesta:2015sga, Evans:2020lbq}. It has also been believed for many years (at least 
since \cite{Hayward:2000ca} though we are confident that the idea goes back further) that apparent horizon jumps
result from the intersection of the spacetime foliation $\Sigma_t$ with a continuous MOTT that may weave backwards and forwards through $\Sigma_t$.  Exact spherically symmetric examples of this behaviour have been known for many 
years \cite{BenDov:2004gh,Booth:2005ng} and there have also been numerical observations in both binary merger
\cite{Schnetter_2006} and strong gravitational wave \cite{Chu:2010yu} spacetimes. 
Very recently it was  
shown that during an axisymmetric merger, such a structure does indeed connect the two initially separate apparent
horizons to the 
final remnant \cite{PhysRevLett.123.171102,PhysRevD.100.084044,pook-kolb2020I,pook-kolb2020II}.

%
%
%
%

The reason that the complicated self-intersecting \MOTSs that make up this MOTT 
were not seen in earlier studies is that standard apparent horizon finders \cite{Thornburg:2006zb} were not 
capable of finding such structures:   those finders assumed that all \MOTSs are star shaped with respect 
to the origin of the employed coordinate system. However it is now understood that self-intersecting \MOTSs are quite
generic and 
not only restricted to dynamic spacetimes. 
In fact, even in a single fixed Cauchy slice of the Schwarzschild spacetime, it is now known that there 
can exist \MOTSs with an arbitrary number of self-intersections\cite{Booth:2020qhb}.

With this explosion in the number and variety of known examples, it has come to be understood that the possible 
properties of \MOTSs are much more varied than was assumed in many early studies which focused on 
the expected properties of black hole boundaries. In particular a horizon is usually assumed to divide a spacetime
into regions containing outer trapped versus untrapped surfaces: mathematically this translates into an assumption that
such a MOTS is stable.

During a  merger,  all studies so far  have shown that 
the outer and original apparent horizons remain
stable (though the outer horizon does have a brief period with positive inward expansion\cite{pook-kolb2020I}). Further, the dynamical apparent horizons that they generate are spacelike and expanding when dynamic (or null and with vanishing expansion in equilbrium).  These are the expected properties. 

However the inner MOTS that splits from the outer apparent horizon is generically unstable  
and the associated MOTT includes timelike, spacelike and null sections. This signature can even vary over 
an individual MOTS as can the sign of the inward null expansion. 
Similarly the MOTTs generated by the exotic new Schwarzschild \MOTSs are unstable and the associated MOTTs 
also have varying signatures. Further in both axisymmetric merger and pure Schwarzschild spacetimes there are now known to be 
self-intersecting \MOTSs.

%

All of this suggests that if we wish to understand the internal dynamics of a black hole merger and discover the ultimate
fate of the original apparent horizons, then we must approach the study without pre-conceived ideas of how it should 
happen.  We need tools that can identify, evolve and characterize 
unexpected \MOTSs. This paper introduces such tools and in the sequel\cite{PaperII} we apply them to help resolve
the full evolution of the apparent horizons and associated MOTTs during an  axisymmetric merger.

%
%
%
%

This paper is organized in the following way. In Section ~\ref{sec:basics} we fix notation 
and recall the necessary mathematics that describes \MOTSs, including the stability
operator. Section \ref{sec:initial} then introduces a new method for finding 
axisymmetric \MOTSs. This is a shooting method that generates curves that can 
be rotated either  into a full MOTS or MOTOS (marginally outer trapped open surface). We dub these curves MOTSodesics and demonstrate
the finder by identifying many previously unknown \MOTSs in  Brill-Lindquist 
binary black hole initial data.
Inspired by those examples, Section \ref{sec:AppNearby} examines the behaviour of 
nearby MOTSodesics by deriving the equations of MOTSodesic deviation in analogy with the 
better known geodesics and geodesic deviation. For these curves the MOTS-stability operator replaces
the geodesic Jacobi operator. We show that the  stability characterization of a MOTS provides us with information about 
the behaviour of nearby MOTSodesics.  
 Section \ref{sec:conclusions} summarizes the work and 
looks forward to \cite{PaperII}. 

\section{Basic Notions}
\label{sec:basics}

In this section we review the basic geometric tools used to characterize and study apparent horizons and other MOTS.

\subsection{Marginally outer trapped surfaces}
Let $(\MM, g_{\alpha \beta}, \nabla_{\alpha})$ be a smooth four-dimensional spacetime with 
signature $({\scriptstyle -+++})$  and $(\Surf, q_{AB}, \mathcal{D}_A)$ be a smooth two-dimensional spacelike surface in that spacetime, where the Greek 
versus capital Latin indices are used to indicate in which manifold a quantity lives. 
The metric on $\Surf$ is the pullback of the full four-metric
\begin{align}
q_{AB} = e_A^\alpha e_B^\beta g_{\alpha \beta} \, , 
\end{align}
where $e_A^\alpha$ is the pull-back operator from $\MM$ to $\Surf$. In coordinates, if $\{x^\alpha \}$ are coordinates on $\MM$
and $\Surf$ is parameterized as  $x^\alpha = x^\alpha (y^A)$, then
$
e_A^\alpha = \frac{\partial x^\alpha}{\partial y^A}
$.

%
%
%
%
Let $\ell^\pm$ be two linearly independent future pointing null normals to $\Surf$ that are cross-normalized so that
$\ell^+ \cdot  \ell^- = -1$.
The extrinsic curvatures associated with these normals can be broken up into  
trace and trace-free components as the expansion $\Theta_\pm$ and shear $\sigma^\pm_{AB}$
by
\begin{equation}
k_{AB}^\pm := e_A^\alpha e_B^\beta \nabla_\alpha \ell^{\pm}_\beta = \frac{1}{2} \Theta_{\pm} q_{AB} + \sigma^\pm_{AB} \;  . 
\label{eq:expansions}
\end{equation}
Directly $\Theta_\pm = q^{\alpha \beta} \nabla_\alpha \ell_\beta^\pm$ where 
\begin{align}
q^{\alpha \beta} = e^\alpha_A e^\beta_B q^{AB} = g^{\alpha \beta} + \ell^{\alpha}_+ \ell^{\beta}_-  + \ell^{\alpha}_- \ell^{\beta}_+ \; . 
\end{align}

%
%
%
%
Within these restrictions, there is  still the freedom to rescale $\ell^\pm$ by a positive function
$f > 0$ via 
\begin{align}
\ell^+ \to \tilde\ell^+ = f\ell^+ \; \; \mbox{and} \; \; 
\ell^- \to \tilde \ell^- = \frac{1}{f} \ell^- \; . \label{rescalings}
\end{align}
Such a scaling leaves the signs of the expansions, and in particular the
conditions $\Theta_{\pm} = 0$, invariant.
If $\Surf$ is closed and orientable, we shall call $\ell^+$ the {\em outgoing}
and $\ell^-$ the {\em ingoing} null normal and $\Theta_{+/-}$
 the outgoing/ingoing expansions\footnote{For simple \MOTSs, 
like $r=2m$ in Schwarzschild, the meaning of ``outgoing'' and ``ingoing'' is clear. However
for the much more complicated twisting and often self-intersecting surfaces that we 
shall consider in this paper, these notions are not always so obvious. For surfaces that intersect
the $z$-axis, ``outgoing'' and ``ingoing'' will be used to refer to the normal directions that
are respectively outgoing or ingoing along that axis and
then consistently propagated everywhere else along the surface. }.

A {\em trapped surface} is a closed surface $\Surf$ with strictly negative
outgoing and ingoing expansion, $\Theta_{\pm}<0$.
If $\Theta_{+} = 0$ with no restriction on $\Theta_{-}$,
then $\Surf$ is called a {\em marginally outer trapped surface} (MOTS).
Similarly, as in \cite{Booth:2020qhb},
we shall call an open surface $\Surf$ with one of the expansions
vanishing a {\em marginally outer trapped open surface} (MOTOS)
and, by convention, call the vanishing expansion $\Theta_{+}$.

A three-surface foliated by \MOTSs is a \emph{marginally outer trapped tube} (MOTT) \cite{Ashtekar:2003hk,Ashtekar:2002ag}. Other references have referred
to these (or very similar) structures as trapping horizons \cite{Hayward:1993wb}, 
future holographic screens \cite{Bousso:2015mqa} or as a generalized dynamical horizon\cite{pook-kolb2020I,pook-kolb2020II}.

\subsection{The MOTS stability operator}
\label{sub:stabilityTheory}

We now consider how \MOTSs may be deformed within a Cauchy slice $\Sigma$. In such a case it is  
natural to write the null normals in terms of the 
unit timelike normal $u^a$ to $\Sigma$ plus the spacelike unit normal $R^a$ to $\Surf$ in $T\Sigma$. From these
we can define gauge-fixed null normals
\begin{align}
\bar{\ell}_+  = \frac{1}{2} (u + R)  \; \; \;  \mbox{and} \; \; \; 
\bar{\ell}_-  =  u - R \nonumber  \; . \label{nullnormals}
\end{align} 

Now for a given initial MOTS $\Surf$, consider a smooth deformation
$\Surf_\nu$ such that $\Surf_0 = \Surf$. 
%
Then the unit normal vector $R^a$ to $\Surf$ naturally extends to a field $R^a$ over the region covered by $\Surf_\nu$
and we can write the tangent vector to the curves that generate this family of deformations as
\begin{align}
\frac{\partial}{\partial \nu} = \psi R 
\end{align}
for some function $\psi$ (a deformation ``lapse'' function). 
In fact the deformation for $\nu = 0$ is entirely determined by specifying $\psi$ over $\Surf$. 

Extending the gauge-fixing (\ref{nullnormals}) to $\Surf$,  the \emph{MOTS stability operator}
on $\Surf$ is defined as the derivative of the gauge-fixed null expansion $\bar{\Theta}_+$ with respect to $\nu$:
\begin{equation}\label{eq:stabilityDef}
    \bar{L}_\Sigma \psi := \delta_{\psi R^a} \bar{\Theta}_{+} 
        := \frac{\partial}{\partial\nu}\Big|_{\nu=0}   \bar{\Theta}_{+}^\nu \; . 
\end{equation}
%

Continuing to use bars to indicate quantities evaluated with the gauge-fixed null normals (\ref{nullnormals}), 
this can be shown to take the form(e.g. \cite{Newman_1987,Andersson:2005gq,Andersson:2007fh})\footnote{Note
that here some numerical factors of $2$ are different from the original derivations due to different scalings of the null 
normals. This is irrelevant to our conclusions.}:
\begin{align}
\bar{L}_\Sigma \psi  =  - \bar{\triangle} \psi
+ \left(\frac{1}{2} \mathcal{R}  - 2|\bar{\sigma}_+|^2 -  2\bar{G}_{\scriptscriptstyle{++}} - \bar{G}_{\scriptscriptstyle{+-}} \right) \psi \label{stab} \; .  
\end{align}
with
\begin{align}
\bar{\triangle} \psi = (\mathcal{D}_A - \bar{\omega}_A)(\mathcal{D}^A - \bar{\omega}^A) \psi \label{triangle}
\end{align}
and
\begin{align} 
\bar{\omega}_A := e_A^a K_{ab} R^b = -e_A^\alpha \bar{\ell}^-_\beta \nabla_\alpha \bar{\ell}_+^\beta 
\end{align} 
is the connection on its normal bundle, with $K_{ab}$ the extrinsic curvature of $\Sigma$ in $M$. 
The other quantities are $\mathcal{R}$ the Ricci scalar on $\Surf$ , $ |\sigma_+|^2 = \bar{\sigma}^+_{AB} \bar{\sigma}^{AB}_+$  the square of the  shear, and
$G_{\scriptscriptstyle{++}} = G_{\alpha \beta} \bar{\ell}_+^\alpha \bar{\ell}_+^\beta$ and
$G_{\scriptscriptstyle{+-}} = G_{\alpha \beta} \bar{\ell}_+^\alpha \bar{\ell}_-^\beta$ determined by the null normals and Einstein tensor. 

$L_\Sigma$ is a second order, linear elliptic operator with a discrete spectrum of eigenvalues, which is self-adjoint if $\bar{\omega}_A = 0$. 
However even if $\bar{\omega}_A \neq 0$, its principal eigenvalue $\lambda_o$ (the eigenvalue with the smallest real part)
is always real. Following \cite{Andersson:2005gq}, a MOTS $\Surf$ is said to be \emph{strictly stable} if 
$\lambda_o > 0$, \emph{stable} if $\lambda_o \geq 0$ and \emph{unstable} if $\lambda_o < 0$. 

As noted in the introduction we use \emph{apparent horizon} as a synonym for a stable MOTS. This 
makes our definition foliation dependent: $R$ is the normal to $\Surf$ in $\Sigma$. However this is a feature
of the definition rather than a bug.  A strictly stable MOTS is one for which 
\begin{align}
\delta_{\psi R} \bar{\Theta}_+ > 0
\end{align}
for all $\psi > 0$. Hence, in the slice all possible deformations outwards result in  outer untrapped surfaces while all possible inward 
deformation result in outer trapped surfaces. Hence a strictly stable MOTS is a  boundary between
trapped and untrapped regions in $\Sigma_t$. 

Unstable \MOTSs are not boundaries of this type: the existence of both 
positive and negative eigenvalues means that the various possible deformations can result in different types of 
surfaces.  Stable but not strictly stable 
MOTS ($\lambda_o = 0$) are a transition between these cases. This interpretation of the deformation operator 
as indicating whether or not there are fully trapped surfaces nearby, precedes the work on the stability operator   \cite{Newman_1987,Hayward:1993wb}\footnote{There is work to define stability in a foliation-invariant way 
(e.g. \cite{Hayward:1993wb,Ashtekar:2003hk,Booth:2006bn}) however that comes with its own complications. 
Here we will be content with foliation dependence.}.

If $L_\Sigma$ has no vanishing eigenvalues then it is invertible. Then it was shown in  
\cite{Andersson:2005gq, Andersson:2007fh} that that MOTS may be 
locally evolved into a MOTT. In particular this is true for a strictly stable MOTS. This time evolution is important  
in the sequel paper and we will return to it there. However for now we are mainly interested in what 
$L_\Sigma$ can tell us about the near-$\Surf$ geometry of  $\Sigma$. 

To that end we consider the general eigenvalue problem
\begin{align}
\bar{L}_\Sigma \psi = \lambda \psi \; . 
\end{align}
As we shall see, important geometric information is conveyed by the number of negative eigenvalues as well as the 
number and location of the zeros of the eigenfunctions. Hence one might be concerned about effects of our 
gauge-fixing. Happily, it has been shown\cite{Jaramillo_2015} that under the  rescaling (\ref{rescalings})
\begin{align}
\bar{L}_\Sigma \psi  = \lambda \psi \; \Longrightarrow  {L}_\Sigma \left( f\psi \right)  = \lambda \left( f \psi \right) 
\end{align}
thanks to the connection terms in (\ref{triangle}). That is, the eigenvalue spectrum is invariant under rescalings of the 
null normals. Further if $\psi$ is an eigenfunction of $\bar{L}_\Sigma$ then $f\psi$ is an eigenfunction of the rescaled
$L_\Sigma$ and hence, with $f\neq0$, it will have the same zeros. These are the properties in which we are interested and so it is 
sufficient to work in the convenient gauge defined by (\ref{nullnormals}). 

\subsection{$\Lslice$ for vacuum, non-spinning, axisymmetry}

The connection $\omega_A$ is closely related to  the angular momentum associated with a black hole (e.g. \cite{Brown:1992br,Hayward:1993wb,Ashtekar:2002ag}) and if it vanishes there can be no angular momentum. We 
will call such a case non-spinning and here were are mainly interested in non-spinning, axisymmetric MOTS in 
vacuum spacetimes. Then
\begin{equation}\label{eq:stability}
    \Lslice \psi = \left( -\Delta_\Surf + \frac{1}{2}\mathcal{R} - 2 |\sigma_+|^2 \right) \psi \; . 
\end{equation}
where we have dropped the overbars since  we have seen that the properties in which we are interested are gauge invariant. 
In the absence of the first order derivatives coming from a non-zero $\omega_A$,  
this is a self-adjoint operator and so the eigenvalue 
spectra that we encounter in this work will always be purely real.

An additional simplification is possible in manifest axisymmetry. 
Consider coordinates $(\theta,\phi)$ on $\Surf$, where $\phi$ is the
coordinate along the orbits of the Killing field $\varphi^a$ which preserves
the induced 2-metric $q_{ab}$ on $\Surf$ and which vanishes precisely at the
two poles.
We can choose $\varphi^A$ such that $\phi \in [0,2\pi)$.
For the following construction, $\theta$ can be any coordinate on $\Surf$
orthogonal to $\phi$.
For definiteness, consider here $\cos\theta = \zeta$, where the invariant
angle $\zeta$ is constructed as in \cite{Ashtekar:2004gp}.
To simplify the eigenvalue problem $\Lslice \psi = \lambda \psi$, we make the
ansatz
\begin{equation}\label{eq:psiAnsatz}
    \psi(\theta,\phi) = \sum_{m=-\infty}^{\infty} \psi_m(\theta) e^{im\phi}
    \,.
\end{equation}
For any fixed value of $m\in\mathbb{Z}$, we can then solve the remaining
one-dimensional problem
\begin{equation}\label{eq:stability1D}
    \Lslice^m \psi_m := (\Lslice + m^2 q^{\phi\phi}) \psi_m = \lambda \psi_m
\end{equation}
and label the resulting eigenvalues with $\lambda_{l,m}$
(c.f. \cite{pook-kolb2020dynamical}).
The index $l$ is assigned in ascending order to the eigenvalues
of Eq.~\eqref{eq:stability1D}, starting, by convention, with $l=|m|$.
Henceforth, the principal eigenvalue is denoted $\lambda_{0,0}$ and we shall write
$\lambda_{l} := \lambda_{l,0}$.

\section{MOTSodesics}
\label{sec:initial}

As was demonstrated in 
\cite{PhysRevLett.123.171102,PhysRevD.100.084044,pook-kolb2020I,pook-kolb2020II}, during
black hole mergers there are exotic \MOTSs that cannot be found with traditional apparent horizon 
finders. However
the new methods introduced in those papers are still somewhat restricted as one needs an idea of 
the kind of geometries that might be possible before being able to find such \MOTSs. Subsequently,
unexpected and even more exotic MOTS geometries were found in pure Schwarzschild \cite{Booth:2020qhb}. Here we introduce a general shooting method that can 
be used to identify axisymmetric \MOTSs with arbitrarily complicated geometries in arbitrary axisymmetric
spacetimes (both exact and numerical). 

This method is related to but better in every way than that used in \cite{Booth:2020qhb}: it is faster, more intuitive and much easier to use. In that 
paper the \MOTSs were parameterized in terms of $\theta$ or $r$ and, for each, a single second order ODE was used to solve for the 
surfaces. However it was necessary to switch back and forth between the equations as a surface became tangent to one or the other of 
these coordinates. For complicated geometries that could mean having to piece together tens of integrations. The new method 
instead uses an arclength parameterization (along the curve that rotates to become the full MOTS) and rewrites the equations as a pair of coupled second order ODEs. Then only a single 
coordinate parameter is needed to cover any of the studied geometries. Further one can leverage physical intuition about 
particles moving in potentials to understand the resulting curves. 

In fact, this method is nearly the same as  some of the original procedures used to find apparent horizons\cite{CADEZ1974449,PhysRevD.16.1609}. The main 
difference is that our method is generalized to find \MOTSs in any axisymmetric spacetimes (including dynamic numerical solutions), applies an arc-length parameterization and is implemented on a
modern computer with modern software! This last point in particular makes it much easier to explore the parameter space of \MOTSs. What was difficult
in the 1970s is now nearly trivial. For exact solutions, the equations are easily solvable with standard mathematical packages while for general numerical data the algorithm has been implemented in 
\cite{pook_kolb_daniel_2021_4687700}.

The basic idea of the method is to rewrite the $\Theta_+ = 0$ condition for an axisymmetric surface in three-dimensional space into a pair 
of coupled ODEs for the generating curve.  For reasons that will become obvious we will refer to such a curve
as a MOTSodesic. We begin by considering how a half-plane rotates into three-dimensions.

\subsection{Rotating the half-plane} 
Consider the spacelike half-plane $\{ \bar{\Sigma}, \bar{h}_{ab}, \bar{D}_{a} \} $ with lower-case Latin indices running over the coordinates $(\rho,z)$ 
which in turn satisfy $\{ \rho>0, - \infty < z < \infty\}$. We rotate $\bar\Sigma$ into 
an axisymmetric three-surface $\{\Sigma,h_{ij},D_i \}$ so that each point becomes a circle. To do this we specify the circumferential radius $R(\rho,z)$ of each point: $(\rho,z)$ maps into a circle with circumference $2 \pi R(\rho,z)$. \\

\textit{Cylindrical-type coordinates: }
In cylindrical-type coordinates $(\rho,z, \phi)$ the three-metric on $\Sigma$ is then 
\begin{align}
h^{(\rho,\phi,z)}_{ij} 
= \left[ \begin{array}{ccc}
\bar{h}_{\rho \rho}  & 0 & \bar{h}_{\rho z} \\
0 & R^2  & 0 \\
\bar{h}_{\rho z} & 0 &   \bar{h}_{zz}
\end{array} \right] \label{hK}
\end{align}
with rotational Killing vector field $\varphi = \frac{\partial}{\partial \phi}$ and 
\begin{align}
\lim_{\rho \rightarrow 0}R(\rho,z) = 0 \; . 
\end{align}
As $\bar{h}_{ab} = h_{ab}$ we now drop the bars when referring to components of the metrics. 

We assume that the surfaces of constant $z$ have been constructed to be conical singularity-free and perpendicular to the $z$-axis. That is: 
\begin{align}
\lim_{\rho \rightarrow 0}\frac{R}{\int_0^\rho \sqrt{h_{\rho \rho}} \mbox{d} \rho} = 1 \; \Longrightarrow \; \lim_{\rho \rightarrow 0}  R_\rho = \lim_{\rho \rightarrow 0}\sqrt{h_{\rho \rho}} 
\label{R_p}
\end{align}
and
\begin{align}
\lim_{\rho \rightarrow 0}\left(   \frac{\partial}{\partial \rho} \cdot \frac{\partial}{\partial z} \right) = 0 \; \Longrightarrow \; \lim_{\rho \rightarrow 0} h_{\rho z} (\rho,z) = 0 \; . 
\end{align}

If we consider $\Sigma$ to be embedded in a similarly symmetric full spacetime $\{M,g_{\alpha \beta}, \nabla_\alpha \}$ and that it be non-spinning ($\omega_A = 0$), the extrinsic curvature of $\Sigma$ in $M$ takes the form
\begin{align}
K^{(\rho,\phi,z)}_{ij} 
= \left[ \begin{array}{ccc}
K_{\rho \rho}  & 0 & K_{\rho z} \\
0 & \Kpp  & 0 \\
K_{\rho z} & 0 &   K_{zz}
\end{array} \right] \label{K} \; . 
\end{align} 
The $R^2$ dependence of the $K_{\phi \phi}$ term is implied by the requirement that $K^i_i$ not diverge. The vanishing $(\rho, \phi)$ and $(z, \phi)$ terms enforce the non-spinning condition.

\textit{Cartesian-type coordinates:}
Alternatively in Cartesian-type coordinates $x = \rho \cos \phi$, $y = \rho \sin \phi$ and $z=z$, the three-metric takes the form
%
%
 \begin{equation}
h^{(x,y,z)}_{ij} \! = \! \left[ \begin{array}{ccc} 
 \frac{x^2 \rho^2  h_{\rho \rho} + y^2 R^2 }{\rho^4}   &   \frac{xy (\rho^2 h_{\rho \rho} -R^2)}{\rho^4}  &  \frac{x}{\rho} h_{\rho z}\\
 \frac{xy (\rho^2 h_{\rho \rho} -R^2)}{\rho^4} &  \frac{y^2 \rho^2  h_{\rho \rho} + x^2 R^2 }{\rho^4} &  \frac{y}{\rho} h_{\rho z}  \\
 \frac{x}{\rho} h_{\rho z}  & \frac{y}{\rho} h_{\rho z} & h_{zz} \\
\end{array} \right] \label{hxyz}
\end{equation}
where  $\rho^2 = x^2 + y^2$.  In this case $\bar{\Sigma}$ is the half-plane $y=0$ and $\rho > 0$ and the rotational Killing vector field 
$\frac{\partial}{\partial \phi} = -y \frac{\partial}{\partial x} + x \frac{\partial}{\partial y}$. 
 

In these coordinates the extrinsic curvature (\ref{K}) becomes
 \begin{equation}
K^{(x,y,z)}_{ij} \! = \! \left[ \begin{array}{ccc} 
 \frac{x^2 \rho^2  K_{\rho \rho} + y^2 \Kpp }{\rho^4} &   \frac{xy (\rho^2 K_{\rho \rho} -\Kpp)}{\rho^4}&  \frac{x}{\rho} K_{\rho z}\\
 \frac{xy (\rho^2 K_{\rho \rho} -\Kpp)}{\rho^4}& \frac{y^2 \rho^2  K_{\rho \rho} + x^2 \Kpp }{\rho^4} &  \frac{y}{\rho} K_{\rho z}  \\
 \frac{x}{\rho} K_{\rho z}  & \frac{y}{\rho} K_{\rho z} & K_{zz} \\
\end{array} \right] \: .  \label{Kxyz}
\end{equation}

\subsection{Curve to two-surface}
\begin{figure}
\includegraphics[scale=0.6]{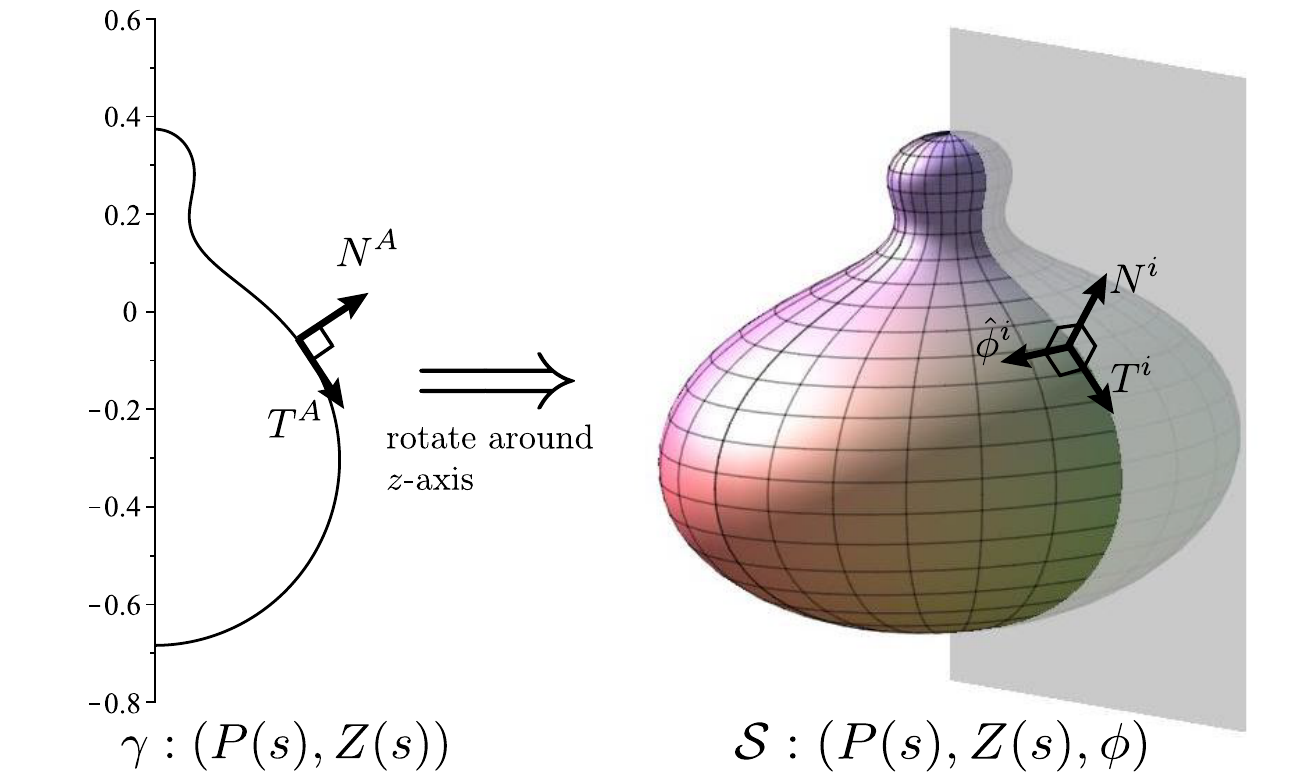}
\caption{A curve $\gamma$ that rotates into a surface $\Surf$. In the text we see how $  \Theta_{\pm} = k_N + k_u$ gives rise to second order equations of motion for $\gamma$ in the $(\rho,z)$ plane. The two-dimensional tangent vector $T^A$ and normal vector $N^A$ to $\gamma$ push-forward to become three-dimensional vectors $T^i$ and $N^i$.  }
\label{TangentNormal}
\end{figure}
Now consider a curve
\begin{equation}
\gamma: (\rho,z) = (P(s),Z(s))
\end{equation}
in $\bar{\Sigma}$ parameterized by an arclength parameter $s$. Then denoting derivatives with respect to $s$ by dots, this has unit-length tangent vector
\begin{align}
T = \dot{P} \frac{\partial}{\partial \rho} + \dot{Z}  \frac{\partial}{\partial z} \; ,
\end{align}
so that 
\begin{align}
h_{ab} T^a T^b = 1 \; . \label{constraint}
\end{align}
The left-hand normal to $\gamma$ is 
\begin{align}
&\tilde{N} =\sqrt{\bar{h}}\left(- \dot{Z} \mbox{d} \rho + \dot{P} \mbox{d} z \right) \; \Longleftrightarrow \; \\
& N = \frac{1}{ \sqrt{\bar{h}}} \left( -( h_{21} \dot{P} + h_{22} \dot {Z}) \frac{\partial}{\partial \rho} +  ( h_{11} \dot{P} +  h_{12} \dot {Z}) \frac{\partial}{\partial z}  \right) \nonumber
\end{align}
where $\bar{h} = \det (\bar{h}_{ab})$ and the acceleration of the curve is given by 
\begin{equation}
T^a \bar{D}_a T^b = \kappa N^b \, , \label{TDT}
\end{equation}
for the signed curvature 
\begin{equation} 
\kappa = N_b T^a \bar{D}_a T^b \;  .  \label{kappa}
\end{equation} 

Next under the rotation that turns $\bar{\Sigma}$ into $\Sigma$, $\gamma$ becomes a 
two-surface $\{\Surf,\tq_{AB}, \mathcal{D}_A\}$ with the indices running over coordinates $(s, \phi)$. 
With $s$ as arclength, the induced metric on $\Surf$ is 
\begin{equation}
{q}_{AB} = 
\left[ \begin{array}{cc}
1 & 0 \\
0 & R^2 
\end{array}
\right] \, . \label{tq}
\end{equation}

In $\Sigma$ we can write the push-forward of the inverse two-metric $\tilde{q}^{AB}$ as
\begin{equation}
{q}^{ij} = T^i T^j + \hat{\phi}^i \hat{\phi}^j 
\end{equation}
where
\begin{equation}
\hat{\phi}^i = \frac{1}{R} \frac{\partial}{\partial \phi} \, . \label{phihat} 
\end{equation}
We can then use that to rewrite the trace of the extrinsic curvature of $\Surf$ in $\Sigma$:
\begin{align}
k_N = {q}^{ij} D_i N_j 
\end{align}
as
\begin{equation}
k_N = \left( T^i T^j + \hat{\phi}^i \hat{\phi}^j \right) D_i N_j = -N_b (T^a \bar{D}_a T^b) +  \hat{\phi}^i \hat{\phi}^j  D_i N_j \, . 
\end{equation}
Equivalently, using the metric (\ref{hK}) to calculate the second term
\begin{equation}
k_N = -\kappa +  N^a \bar{D}_a (\ln R ) \; . 
\end{equation}

Meanwhile from (\ref{K}) the extrinsic curvature of $\Sigma$ can be written as
\begin{align}
K_{ij} = k^u_{ij} + k_{\phi \phi} \hat{\phi}_i \hat{\phi}_j \label{Kex}
\end{align}
where $k^u_{ij} = q_i^{\phantom{i} k} q_j^{\phantom{j} l} K_{ij}$. The non-spinning condition means that there are no $q_i^{\phantom{i} k} \hat{\phi}_j$ ``cross'' terms. Thus 
the trace of the extrinsic curvature of $\Surf$ with respect to the timelike normal $u$ to $\Sigma$ is 
\begin{align}
k_u = \tq^{ij} K_{ij} = k^u_{ab} T^a T^b + k_{\phi \phi} \, , 
\end{align}
though for compactness we will usually continue to write this as $k_u$ in the following expressions. 

The null normals $\ell^{\pm}$ 
for left-hand/right-hand normals respectively) then have expansion 
\begin{align}
  2 \Theta_{+} &=  k_u  + (-N_b (T^a \bar{D}_a T^b) +  N^a \bar{D}_a (\ln R )  ) \\
  \Theta_{-} &=  k_u  - (-N_b (T^a \bar{D}_a T^b) +  N^a \bar{D}_a (\ln R )  ) \nonumber
\end{align}
and so $\gamma$ will rotate to a surface of vanishing null expansion if 
\begin{equation}
N_b (T^a \bar{D}_a T^b) =   N^a \bar{D}_a (\ln R )  \pm k_u  \, . \label{baseEq}
\end{equation} 
For each of the two cases of outward/inward expansion, this is a single differential equation that is second order in $P(s)$ and $Z(s)$. 
Adding in the arc-length constraint (\ref{constraint}) we have a 
pair of coupled differential equations that in principle can be solved for $P(s)$ and $Z(s)$.

It is much easier to solve these equations if we rewrite them as a pair of second order equations: most importantly this 
will avoid awkward sign-changing square root terms that led to errors in \cite{Booth:2017fob} and the complicated (though correct) repeated equation switching of \cite{Booth:2020qhb}. 
This rewriting can be done directly by using the derivative of (\ref{constraint}) to alternately remove $\ddot{Z}$ and then $\ddot{P}$ terms from (\ref{baseEq}). However it is easier to get the equations by matching (\ref{kappa}) and  (\ref{baseEq}) to find
\begin{equation}
\kappa^\pm =   N^a \bar{D}_a (\ln R )  \pm  k_u  \: . \label{kappaC}
\end{equation}
Then (\ref{TDT}), which we will henceforth refer to as the MOTSodesic equation, becomes a pair of second order equations for $\gamma$ in 
$\bar{\Sigma}$:
\begin{align}
T^a \bar{D}_a T^b = \kappa^\pm N^a \label{TDT2}
\end{align} 
which, on expanding out the covariant 
derivatives, become
\begin{align}
\left[  \begin{array}{c}
\ddot{P} \\
\ddot{Z}
\end{array} \right]^a = \dot{T}^a 
= -  \bar{\Gamma}^a_{bc} T^b T^c + \kappa^\pm  N^a \; , \label{MainEq}
\end{align} 
%
where $\bar{\Gamma}^{a}_{bc}$ are the Christoffel symbols in $\bar{\Sigma}$ (or equivalently the $(\rho,z)$ Christoffel symbols in $\Sigma$)
and the dot indicates a regular derivative with respect to $s$. 
The equations  will be fairly complicated for an arbitrary metric and extrinsic curvature but they may still be solved with standard numerical solvers to produce a MOTOS given initial conditions
$P(s_o), Z(s_o), \dot{P}(s_o)$ and $\dot{Z}(s_o)$ for some $s_o$ (typically $0$). 

An important special case of these equations is time-symmetric data: $K_{ij} = 0$. Then $k_u = 0 \Longrightarrow 2\Theta_+ = -\Theta_-$ and  
so they both vanish simultaneously if and only if $k_N =0$. That is, $\Surf$ is a minimal surface in $\Sigma$ and is both marginally outer
and marginally inner trapped. 
%

\subsection{Departing from the $z$-axis}
\label{sec:depart}

There is, however, a complication. 
We mainly use (\ref{MainEq}) to find axisymmetric \MOTSs. Most (though not all: see the toroidal example in  \cite{PaperII})
closed axisymmetric surfaces intersect the $z$-axis.
 Unfortunately this is also where the coordinate system fails and $R \rightarrow 0$. This complicates the calculation of 
the first term in $\kappa^\pm$ from (\ref{kappaC}).

We can sidestep this problem by solving (\ref{MainEq}) in a series expansion near the $z$-axis and then using this
to start evolutions a short distance from the axis. That is we assume
\begin{align}
P &= \phantom{P_0 + } \; \;      P_1 s + \frac{1}{2!}  P_2 s^2 
+ \dots \\
Z &= Z_0  +   Z_1 s + \frac{1}{2!}  Z_2 s^2 
+ \dots  \end{align}
where the requirement that we start from the $z$-axis at $s = 0$ means that the $P_0$ term vanishes. 

$Z_0$ is chosen as initial data. Next, demanding smoothness of $\Surf$ (i.e. no conical singularities) at the $z$-axis
requires 
\begin{align}
 \lim_{s \rightarrow 0}  \dot{R}   = 1 & \; \Longrightarrow \; \lim_{s \rightarrow 0} R_\rho \dot{P} + R_z \dot Z  =  1 \; . 
\end{align}
Since $R_z = 0$ along the $z$-axis (where $R=0$) we find
\begin{align}
P_1 = \lim_{s \rightarrow 0} \frac{1}{R_\rho} \eqz \frac{1}{ \sqrt{h_{\rho \rho}}}   \nonumber
\end{align}
where the overset $z$ indicates that what follows is evaluated for $\rho =0$ on the $z$-axis, and the last equality follows from  (\ref{R_p}). Then from the normalization condition (\ref{constraint}) we also have that 
\begin{align}
Z_1 = 0 \label{perp} \; . 
\end{align}
That is, MOTSodesics can only intersect the $z$-axis at a right angle. 

For many cases this first order expansion will be sufficient, since if we start at small $s_o$ we would then expect the error to be of $O(s_o^2)$. However for those cases 
which require a more accurate expansion we also present the second order term of the expansion (and so have an error of order $O(s_o^3)$). These
follow directly from (\ref{MainEq}). The $\rho$ term is straightforward:
\begin{align}
P_2 \eqz   - \frac{\bar{\Gamma}_{\rho \rho}^\rho}{h_{\rho \rho}}   
\; . 
\end{align}
The $z$-term is more complicated thanks to the expansion of $\kappa^\pm$ for which the first term is of the form  $\frac{0}{0}$ in the limit as $s \rightarrow 0$.  This limit 
is found in Appendix \ref{AppKappa} from which
\begin{align}
\kappa_o^\pm & = \lim_{s \rightarrow 0}  \kappa^{\pm} \eqz \frac{1}{2}   \left(\frac{R_{TN}}{R_T} \pm \left(\frac{K_{\rho \rho}}{h_{\rho \rho}} + k_{\phi \phi} \right) \right) 
\end{align}
where $R_{T} = T^a \! \bar{D}_a  R$ and $R_{TN} = T^a\! N^b\! \bar{D}_a \bar{D}_b R$. Then 
\begin{align}
Z_2 & \eqz    - \frac{\bar{\Gamma}_{\rho \rho}^z}{{h_{\rho \rho}}}  + \frac{\kappa_o}{\sqrt{h_{zz}}}  \; . 
\end{align}
For the cases that we have studied, the second order terms were sufficient to obtain  the necessary
accuracy.

\subsection{New \MOTSs in Brill-Lindquist initial data}
\label{sub:BLinitial}

We now apply this technique to find new \MOTSs in Brill-Lindquist initial black hole data\cite{PhysRev.131.471}. 
This is defined by a conformally flat metric 
\begin{align}
h_{ij} \mbox{d} x^i \mbox{d} x^j = \psi(\rho,z)^4 \left(\mbox{d} \rho^2 + \mbox{d} z^2 + \rho^2 \mbox{d} \phi^2 \right) \label{BLmet}
\end{align}
on $\Sigma$ for which the extrinsic curvature $K_{ij} = 0$ vanishes. Then the diffeomorphism constraint is trivial and the
Hamiltonian constraint reduces to the Euclidean axisymmetric Laplace equation:
\begin{align}
\psi_{\rho \rho} + \psi_{zz} + \frac{\psi_\rho}{\rho} = 0  \; . 
\end{align}
As this is time-symmetric data, the  \MOTSs will be minimal surfaces of (\ref{BLmet}) with $\Theta_+=\Theta_-=0$.

\subsubsection{MOTSodesic equations}

The arclength parameterization condition for this metric can be written as
\begin{align}
\left( \dot{P}^2 + \dot{Z}^2 \right) - \frac{1}{\psi^4} = 0 \, , 
\end{align}
which for intuitive purposes can be usefully interpreted as $\mbox{kinetic energy} +  \mbox{potential energy} = 0$. 

To derive the full equations of motion we note that for this initial data $R = \psi^2 \rho$ and 
\begin{align}
N = - \dot{Z} \frac{\partial}{\partial \rho} + \dot{P} \frac{\partial}{\partial z} \; . 
\end{align}
Then with $k_u = 0$, it is a straightforward calculation to obtain:
\begin{align} 
\kappa^\pm   & =  \bar D_N \ln (R)  =  2  \bar D_N ( \ln \psi )-  \frac{\dot{ Z}}{P} \; . 
\end{align}
For this time symmetric data $\kappa^+ = \kappa^-$ (which also follows from $2\Theta_+ = -\Theta_-$). 
Next
\begin{align}
- \bar\Gamma^a_{bc} T^c T^d & = - 2 \bar{D}_{T} (\ln \psi) T^a +   2  \bar{D}_{N} (\ln \psi) N^a \\
& = \left[ \begin{array}{l}
-2 \dot{P}  \bar{D}_{T} (\ln \psi) - 2 \dot{Z}  \bar{D}_{N} (\ln \psi) \\
-2 \dot{Z}  \bar{D}_{T} (\ln \psi) + 2 \dot{P} \bar{D}_{N} (\ln \psi) 
\end{array} \right]^a \nonumber
\end{align}
where we have abbreviated $T^a \bar{D}_a = \bar{D}_T$ and $N^a \bar{D}_a = \bar{D}_N$. 
Then from (\ref{MainEq}) and a little bit of algebra, the final equations of motion for $\gamma$ are:
\begin{align}
\left[  \begin{array}{c}
\ddot{P} \\
\ddot{Z}
\end{array} \right]^a =  4 \bar{D}^a (\ln \psi)  -\left( \frac{\dot{Z}}{P} \right) N^a    - \left(  6 \bar{D}_T (\ln \psi) \right) T^a \label{BLbasic}\; . 
\end{align}
Note that the first term on the left-hand side is a potential term similar to those found in Newtonian gravity: it generates a coordinate acceleration ``up'' the 
potential. Keeping in mind that $N$ is to the left-hand of $T$, the second term generates a repulsion from the $z$-axis. As $P \rightarrow 0$ this goes to 
infinity (unless $\dot{Z}=0$). The third term causes the curve to ``slow down'' as it moves up the potential or ``speed up'' as it goes down. 

Explicitly these take the form
 \begin{align}
 \label{eq:bleqs}
\ddot{P} &=  \phantom{-;}  \frac{\dot{Z}^2}{P}  + \frac{4 \psi_\rho}{\psi^5}  - \frac{6 \dot{P} (\dot{P} \psi_\rho + \dot{Z} \psi_z ) }{\psi}   \\
\ddot{Z} &=  - \frac{\dot{Z} \dot{P}}{P} + \frac{4 \psi_z}{\psi^5} - \frac{6 \dot{Z} (\dot{P} \psi_\rho + \dot{Z} \psi_z)}{\psi} \; . 
\end{align}
This is the form that we use when numerically integrating MOTOS. As a consistency check notice that these can be combined to find  
$\ddot{P} \dot{P} + \ddot{Z} \dot{Z} = - 4 \dot{\psi}/\psi^5$ as would be expected from the arc length parameterization condition.

\subsubsection{Classes of new marginal surfaces}
\label{sub:classification}

\begin{figure}
\includegraphics{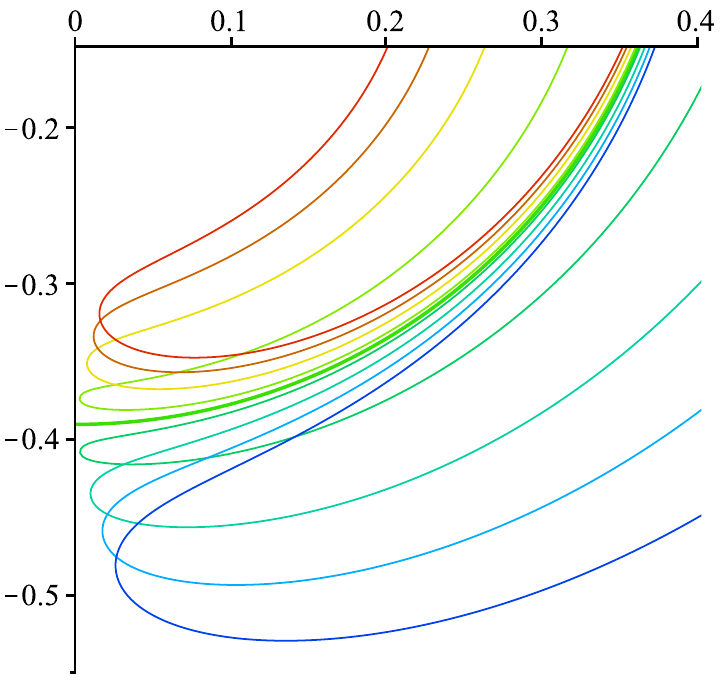}
\caption{The process of using the shooting method to find the inner horizon from FIG.~\ref{fig:BL-standard_horizons}. 
The MOTSodesics are initially parallel when launched perpendicularly from the positive $z$-axis. From
blue to red the initial conditions were $z_o = 0.7297755$, $0.7299755$, $0.7301755$, $0.7303755$, $0.7304755$ 
(central green), $0.7305755$, $0.7307755$, $0.7309755$.   }
\label{ShootingDemo}
\end{figure}

We now consider the numerical solution to these equations. Our procedure is as follows. We consider surfaces that  smoothly intersect the $z$-axis. To ensure this, we use the results of \ref{sec:depart} (or alternatively perform a direct series expansion of \eqref{eq:bleqs} near the axis), including terms up to second-order and use this series expansion as initial conditions, evaluating this series a distance of $10^{-4}$ away from the axis. The integration was performed in Mathematica utilizing the default method available in NDSolve with a working precision of 25. 

Due to the axis-repulsion term in (\ref{BLbasic}) which diverges for $\dot{Z} \neq 0$,
the specification of any particular MOTS requires infinite precision
in the initial conditions. As shown in FIG.~\ref{ShootingDemo}, any deviation from the precise value leads to a strong
repulsion from both the $z$-axis and the MOTS that does reach that axis.
We use the shooting method, adjusting the initial conditions and narrowing the range in which we know the true MOTS must
exist, until the surface can be considered to approximately close.  We consider that  to be an approach to the axis to within 
a distance of about $10^{-6}$ or better. These surfaces are then confirmed to be MOTS using the methods of \cite{pook_kolb_daniel_2021_4687700}. Hence they have been identified by two completely  independent methods.

The binary black hole Brill-Lindquist potential is:
\begin{equation} 
\psi = 1 + \frac{m_1}{2 \sqrt{\rho^2+z^2}} + \frac{m_2}{2 \sqrt{\rho^2 + (z-d)^2}}  \; . 
\end{equation}
This corresponds to a pair of instantaneously stationary black holes ``centred'' at $(0,0)$ and $(0,d)$. Quotation 
marks are employed as those points actually correspond to asymptotic infinities on the other side of Einstein-Rosen
bridges from the usual $\rho^2 + z^2 \rightarrow \infty$ region. Relative to those other infinities the black holes can be 
measured to have ADM masses\footnote{%
In the $\rho^2 + z^2 \rightarrow \infty$ region, the ADM mass is $\MADM = m_1 + m_2$, i.e. it is NOT the sum of the two puncture ADM masses. For a nice discussion of the
full geometry of this initial data see, for example,  \cite{PhysRev.131.471} and \cite{baumgarte_shapiro_2010}. }
$\MADM^i = m_i + m_1m_2/2d$. 

\begin{figure}
	\centering
    \includegraphics[scale=0.4]{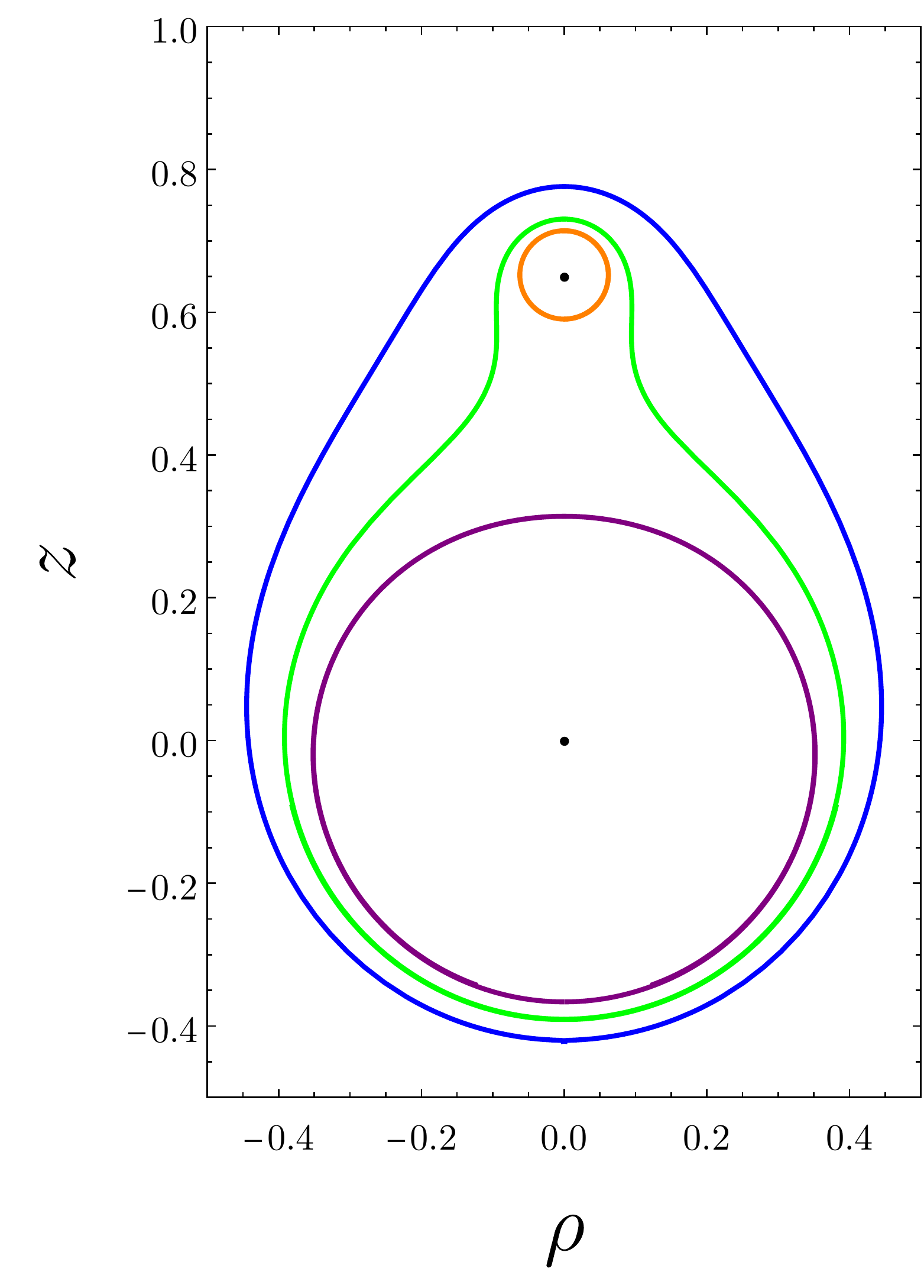}%
    \caption{\label{fig:BL-standard_horizons} Standard MOTS in the BL initial data: the three apparent horizons (the two 
    original plus the common) as well as the unstable inner MOTS (green).} 
\end{figure}

For the purposes of illustration, we take $m_1=0.2$,  $m_2 = 0.8$ and $d = 0.65$. With these values, we can readily reproduce the well-known structure of horizons in this geometry, as shown in FIG.~\ref{fig:BL-standard_horizons}.
In the usual way the outer and individual MOTS are strictly stable in the sense of \ref{sub:stabilityTheory} and so we
call them apparent horizons. However the
inner (green) MOTS is unstable with one negative eigenvalue \cite{pook-kolb:2018igu} and so we do not think of it as a horizon. 

\begin{figure*}
	\centering
	\includegraphics[width=0.3\linewidth]{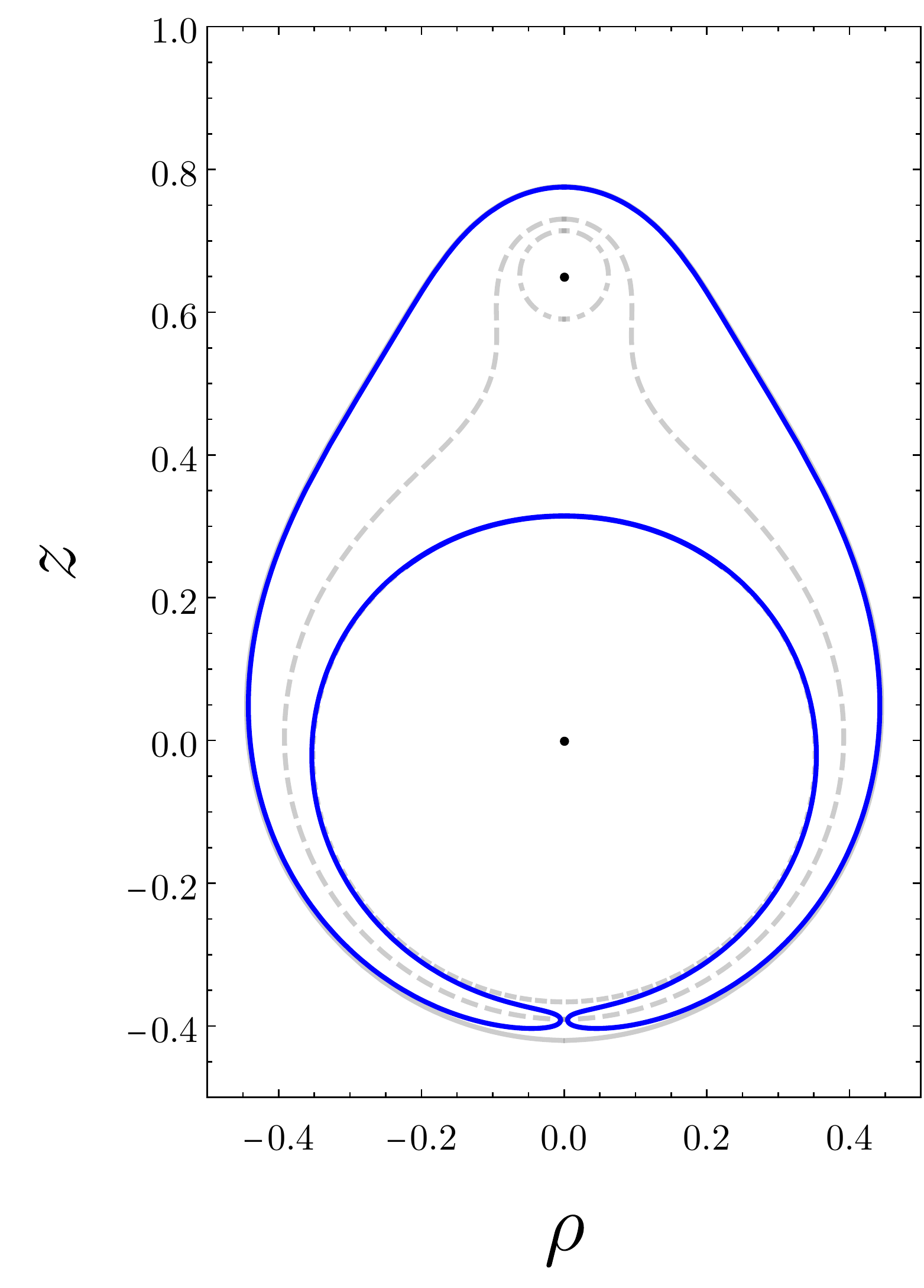}
	\includegraphics[width=0.3\linewidth]{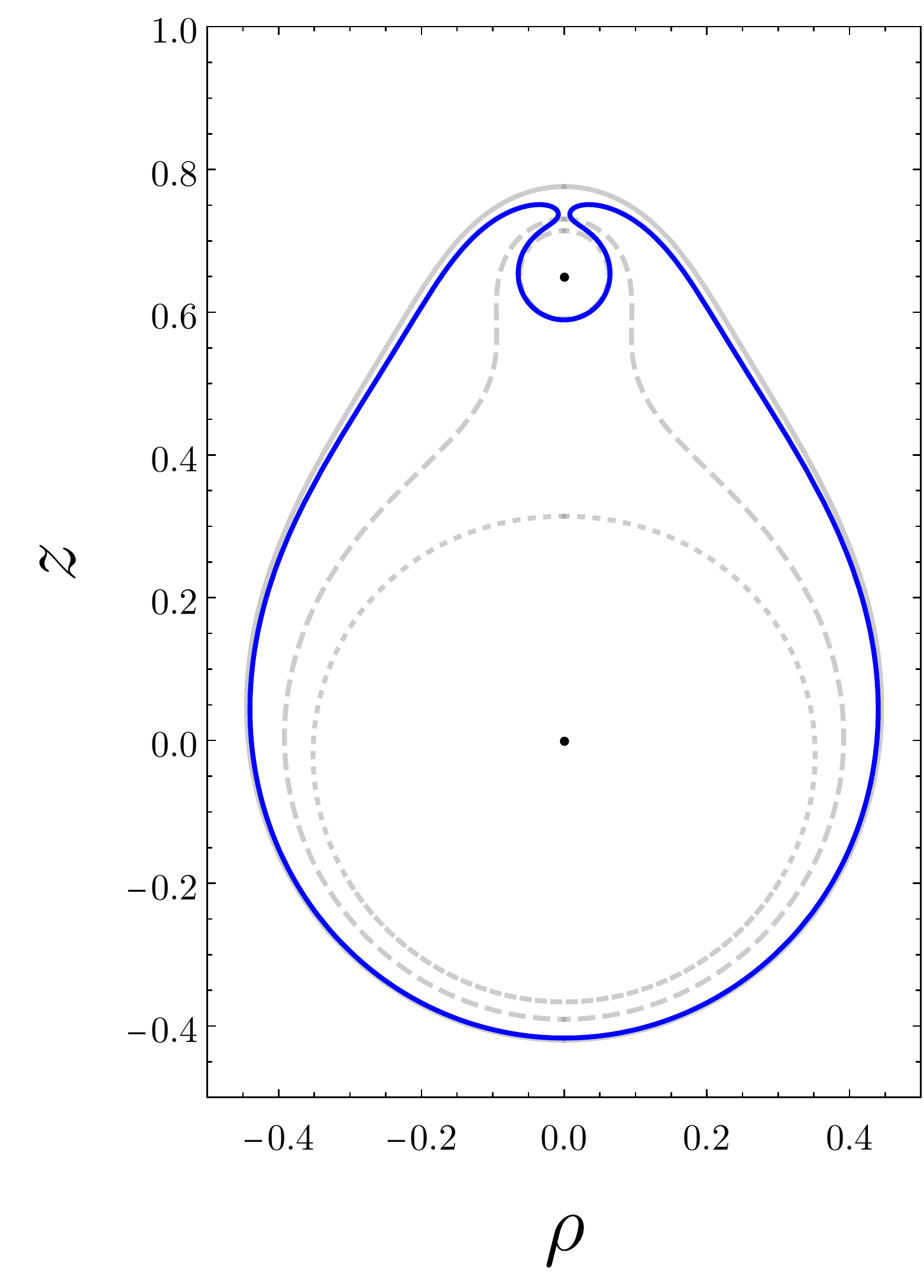}
	\includegraphics[width=0.3\linewidth]{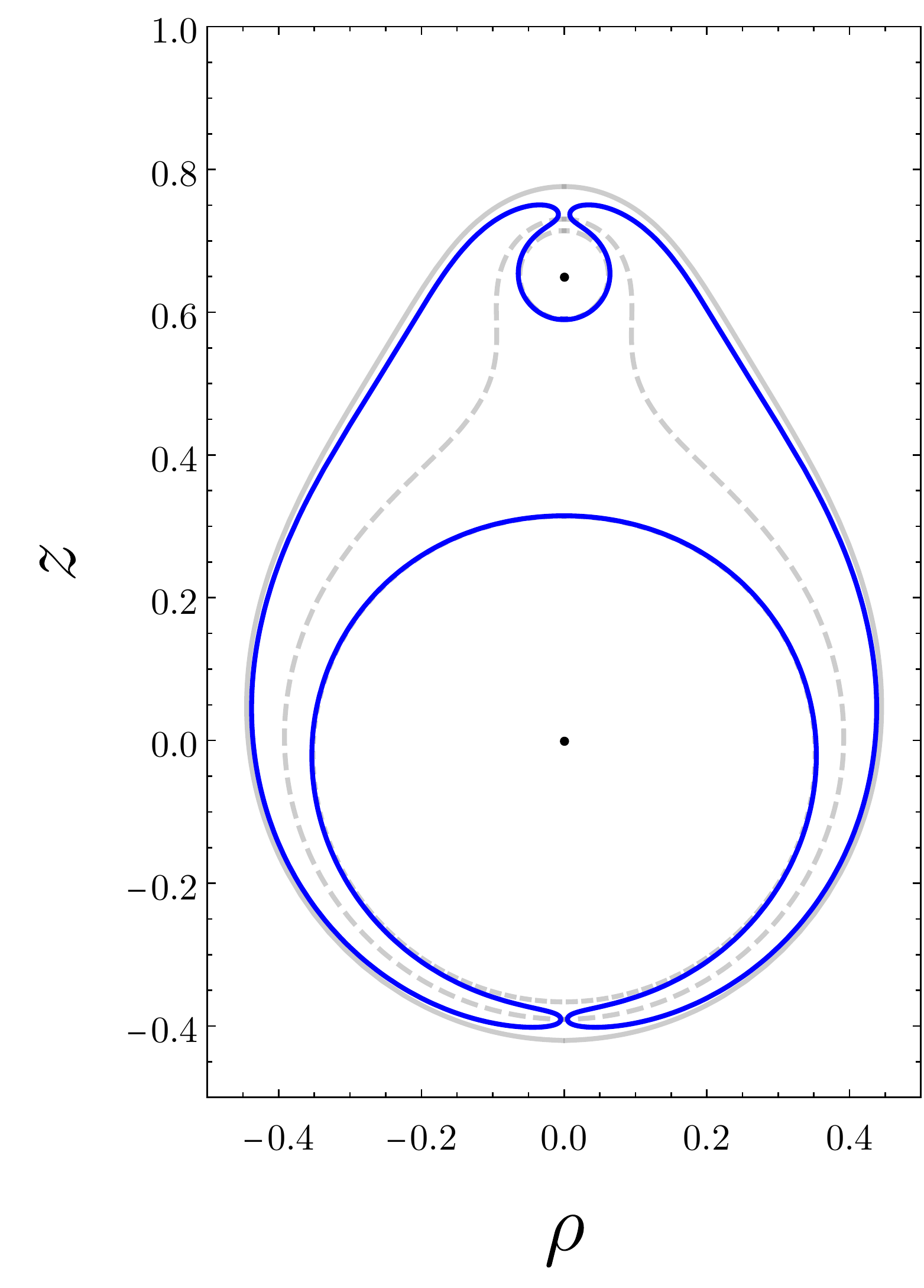}
	\includegraphics[width=0.3\linewidth]{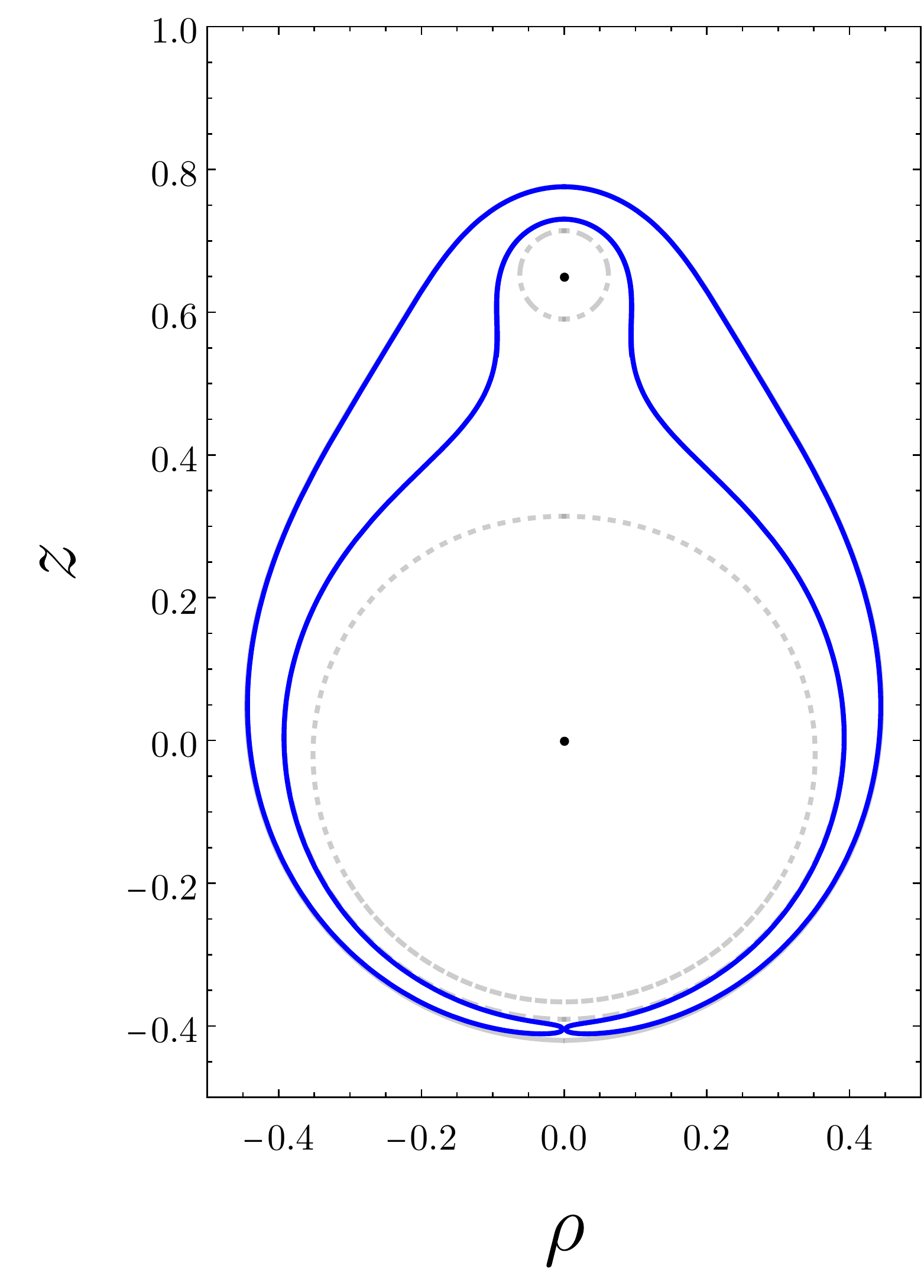}
	\includegraphics[width=0.3\linewidth]{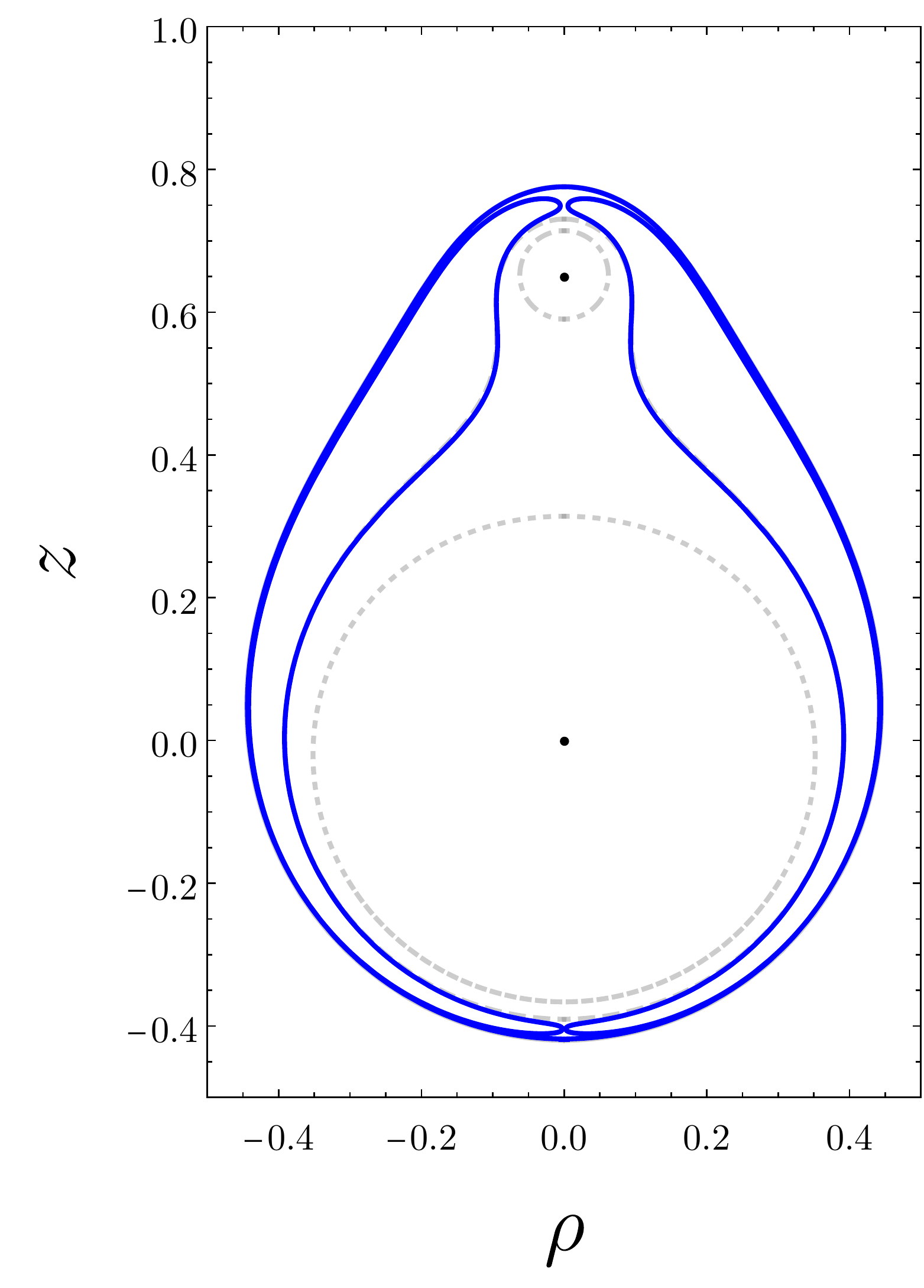}
	\includegraphics[width=0.3\linewidth]{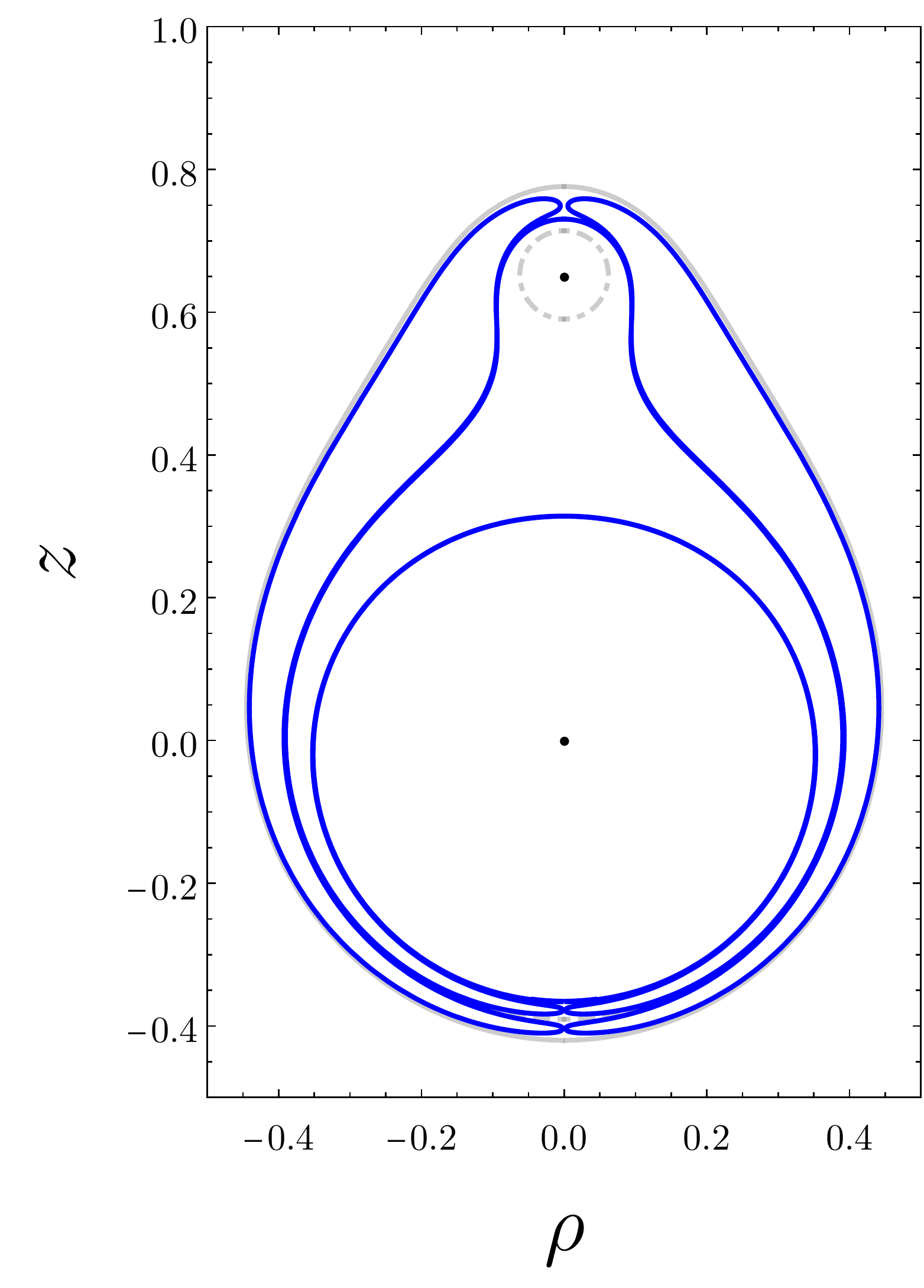}
	\caption{Several examples of new \MOTSs in Brill-Lindquist initial data. In order of left-to-right top-to-bottom, the curves can be reproduced using the method described in the text starting from initial conditions: $z_0 = .775299193$, $z_0 =-.416830101$, $z_0 =.31478148$, $z_0 =.7756759$, $z_0 =.7756696814$, and $z_0 =.730402549931$. As infinite precision is required to exactly specify the initial condition for a MOTS, it should be understood here that there is an uncertainty of $\pm 1$ in the last digit of the initial conditions presented. Furthermore, in each case we have presented initial conditions to the minimum precision that is needed to declare the curves to close.}
	\label{fig:moreMOTS}
\end{figure*}

It is natural to wonder whether additional axisymmetric \MOTSs exist, and a careful search reveals many such surfaces, several of which are shown in Fig.~\ref{fig:moreMOTS}. The additional \MOTSs appear to defeat any simple attempt to classify them, but general features are illuminating. New \MOTSs can be found between the outer apparent horizon and the apparent horizons associated with the  two punctures. No MOTS that we have located extends outside the outer apparent horizon or inside the  two inner apparent horizons. The new \MOTSs can enclose either, both, or neither of the two punctures. 

We find that these surfaces tend to `hug' closely the common apparent horizon, and/or the individual \MOTSs.  
It appears that \MOTSs exist that wrap these surfaces an arbitrary number of times (we have generated examples involving up to ten folds). An equivalent behaviour was seen around  $r=2m$ for pure Schwarzschild in regular 
time-symmetric Schwarzschild coordinates in\cite{Booth:2020qhb}. 
We will return to this behaviour in \ref{sec:return_to_BL} and here just note that  it is related to the time-symmetry of these 
slices. The \MOTSs are minimal surfaces which do not have a distinguished direction of vanishing null normal expansion. 
That is $\Theta_+ = \Theta_- = 0$. Hence a MOTS that is close to one of the horizons can turn around at the $z$-axis,
still be close, and continue to hug. Thus if it starts sufficiently close it will gradually work its way outwards with an 
extra kick each time it approaches the $z$-axis. 

Intuitively \MOTSs can ``orbit'' (though gradually recede from) the stable \MOTSs. However it is also possible to 
jump between orbits. Then the unstable inner MOTS from FIG.~\ref{fig:BL-standard_horizons} can be thought of as the first of these joint orbits. 

Clearly many of these \MOTSs have multiple segments that are nearly ``parallel''. Hence in the next section we study the 
relative evolution of nearby MOTSodesics. Once we have developed those methods we will return to these examples. 

%


\section{MOTSodesic deviation}
\label{sec:AppNearby}

To better understand some of these behaviours let us consider the evolution of MOTSodesics around a central one. 
This is the analogue of calculating the geodesic deviation equation and as for that case can be thought of as a 
linearization of the MOTSodesic equations around a central curve $\gamma_o$. Our main result is to demonstrate  
that this relative evolution is governed by the stability operator. 

\subsection{Via quantities defined in $\bar{\Sigma}$ }
Consider a 
congruence of curves $\gamma_\chi (\lambda)$ where $\chi$ labels the curves and $\lambda$ parameterizes them. 
These live in the half-plane $\{\bar{\Sigma}, \bar{h}_{ab}, \bar{D}_a\}$, as shown in Fig.~\ref{congruence}.  We assume that each rotates
into a MOT(O)S: 
\begin{figure}
\includegraphics{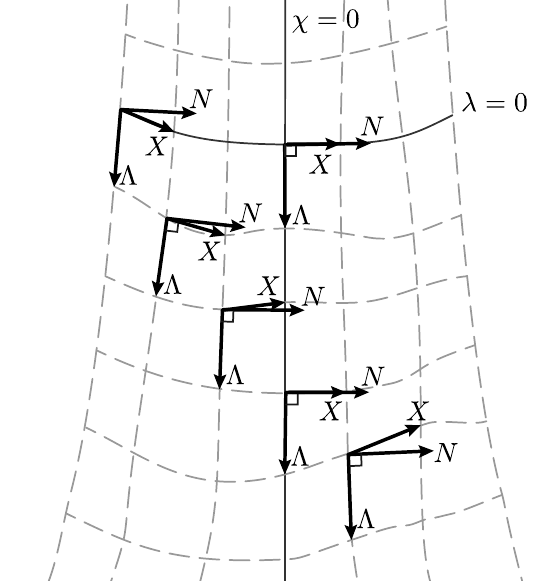}
\caption{A congruence of MOTSodesics  $\gamma_\chi(\lambda)$ in $\bar{\Sigma}$. $\chi$ labels the curves and $\lambda$ is a parameter along the curves. 
The corresponding coordinate vector fields are $X = \frac{\partial}{\partial{\chi}}$ and $\Lambda = \frac{\partial}{\partial \lambda}$.The parameters are chosen so that along $\gamma_0 (\lambda)$, $\lambda$ is the arclength
parameter $s$ (so $\Lambda = T$) and $X \perp T$. However when $\chi \neq 0$ these will not usually hold. }
\label{congruence}
\end{figure}
that is, each curve  satisfies (\ref{TDT})
for unit tangent vector ${T}$. 

As long as the curves don't intersect, the $(\lambda, \chi)$ are a good coordinate system in a neighbourhood of $\gamma = \gamma_0 (\lambda)$. The
coordinate tangent vectors are
\begin{align}
\Lambda = \pdv{\lambda} \; \; \mbox{and} \; \; X = \pdv\chi \; . 
\end{align}
In the standard calculation for geodesic deviation (e.g.\cite{Wald:106274}), $X$ can be understood
as a deviation vector which points to nearby geodesics. The rest of this section can be understood as a modified 
version of  that calculation. 

For some functions $\mu(\lambda, \chi)$, $\alpha (\lambda, \chi)$ and $\beta (\lambda, \chi)$ we can decompose these coordinate vectors  as
\begin{align}
\Lambda =  \mu T \; \; \mbox{and} \; \;  X = \alpha T + \beta N \; , \label{LX}
\end{align}
where, as usual, $N$ is the (left-hand) unit  normal to the $\gamma_\chi$. 
It is always possible to adjust the $\lambda$ between the curves, ($\lambda,\chi) \rightarrow (\tilde{\lambda}(\chi,\lambda),\chi) $, so that: 1) $\lambda = s$ on $\gamma$: that is $\mu(\lambda, 0) = 1$ 
and 2) $X \perp T$ on $\gamma$: that is $\alpha(\lambda, 0) = 0$. However for $\chi \neq 0$, $\lambda$ will generally 
be a non-affine
parameter  and $\alpha \neq 0$.

By (\ref{TDT2}) and the fact that $T$ and $N$ are orthonormal we have
\begin{align}
T^a \bar{D}_a T^b = \kappa N^b \; \; \mbox{and} \; \; T^a \bar{D}_a N^b = -\kappa T^b
\end{align}
while for some function $\nu(\lambda,\chi)$ we similarly have 
\begin{align}
N^a \bar{D}_a T^b = - \nu N^b \; \; \mbox{and} \; \; N^a \bar{D}_a N^b = \nu T^b \; . 
\end{align}
Geometrically $\kappa$ and $\lambda$ define the connection for the $(T,N)$ dyad. 
For notational conciseness we have dropped the $\pm$ superscript from the $\kappa^\pm$ and continue to do this for the rest of the section. 

Since $\Lambda$ and $X$ are coordinate vector fields we have
\begin{align}
X^a \bar{D}_a \Lambda^b = \Lambda^a \bar{D}_a X^b \, , \label{coord}
\end{align}
and so on expanding we find
\begin{align}
\mu' & = \dot{\alpha} - \kappa \mu \beta  \label{dota}\\
 \dot{\beta} & = - \nu \mu \beta \, , \label{dotb} 
\end{align}
where the dot generalizes here to indicate a partial derivative with respect to $\lambda$ and the prime is a 
partial derivative with respect to $\chi$.

Next, still following the example of the geodesic deviation calculation, we use these results to find 
equations of motion for the deviation vector $X$ along $\gamma$:
\begin{align}
A^a   \equiv& \Lambda^a \bar{D}_a (\Lambda^b \bar{D}_b X^c)  \label{E1} \\
  =&  \left( \partial_\lambda [\dot{\alpha} - \kappa \mu \beta] - \kappa \mu [\dot{\beta} + \kappa \mu \alpha] \right) T^a \nonumber \\
& + \left(\partial_\lambda[\dot{\beta} + \kappa \mu \alpha] + \kappa \mu [\dot{\alpha} - \kappa \mu \beta] \right) N^a \nonumber  \; . 
\end{align}
Alternatively we can apply (\ref{coord}) to write 
\begin{align}
A^c   = \Lambda^a \bar{D}_a (X^b \bar{D}_b \Lambda^c) \; , 
\end{align}
which can be expanded and manipulated to find
\begin{align}
A^a =  & \left(\dot{\mu}' - \mu^2 \kappa (\alpha \kappa - \beta \nu) \right) T^a \\
& + \left( (\mu^2 \kappa)' + \dot{\mu} (\alpha \kappa - \beta \nu) - \mu^2  \beta \mathcal{K} \right) N^a \, , \label{E2}
\end{align}
where  $\mathcal{K}= \frac{1}{2} \bar{h}^{ac} \bar{h}^{bd} \bar{\mathcal{R}}_{abcd} $ is the Gauss curvature of $\bar{\Sigma}$.

Applying (\ref{dota}) and (\ref{dotb}) it is straightforward to see that the $T$-terms are the same. However
on matching the $N$-terms we obtain a differential equation for $\beta$:
\begin{align}
\ddot{\beta} + (\mathcal{K} \mu^2 - \kappa^2 \mu ) \beta = \mu^2 \kappa' + \dot{\kappa} \mu \alpha - \nu \dot{\mu} \beta \; . 
\end{align}
Restricting to $\gamma$ where $\mu = 1$ and $\alpha = 0$ this takes the form
\begin{align}
\ddot{\beta} +  (\mathcal{K} + \kappa^2  - N^a \! \bar{D}_a \kappa) \beta \eqg 0 \; .   \label{ddb1}
\end{align}
From (\ref{LX}), $\beta$ should be understood (in the linear regime) as the perpendicular distance from $\gamma_0$ to the nearby $\chi$-geodesic. 

%
%
%
%
%

This is a general expression that would hold for any specification of $\kappa$ . However we have an 
expression (\ref{kappaC}) for it in terms of $R$ and $k_u$. In terms of quantities defined on
$\bar{\Sigma}$ this can be written as:
\begin{align}
\kappa = N^a \bar{D}_a (\ln R) \pm (K - K_{ab} N^a N^b)  \; . 
\end{align}

\subsection{Via quantities defined in $S$ }

While these expressions may be useful when studying particular MOTSodesics, to study their general properties it is
more useful to rewrite the $\beta$-term of (\ref{ddb1}) in terms of quantities defined
on $\Surf$. To do this we make use of the Einstein equations in their constraint form:
\begin{align}
\mathcal{R}^{\Sigma} + K^{ij}K_{ij} - K^2 & = 0 \label{HamCon} \\
D_j K_i^{\phantom{i} j} - D_i K & = 0 \; . \label{DiffCon}
\end{align}

First we decompose $\mathcal{R}^\Sigma$ into quantities defined on $\bar{\Sigma}$ plus the unit 
normal 
$\hat{\phi}_a$.  Using the Gauss-Codazzi equations along with the definition of the Riemann tensor in terms of commuting derivatives, one can show
(this is a standard expression for hypersurfaces derived, for example, as Eqn.~(3.43) in \cite{poisson_2004}):
\begin{align}
\mathcal{R}^\Sigma = 2 \mathcal{K} \!+\!  (k_{\hat{\phi}}^2 - k_{\hat{\phi}}^{ab}  k^{\hat{\phi}}_{ab}) \!+\! 2 D_i \!\left(\hat{\phi}^j D_j \hat{\phi}^i \! - \! \hat{\phi}^i D_j \hat{\phi}^j  \right) 
\end{align}
where $k^{\hat{\phi}}_{ab} = e_a^i e_b^j D_i \hat{\phi}_j$. But $\pdv{\phi}$ is a Killing vector field,  so $k^{\hat{\phi}}_{ab} = 0$. Next, using (\ref{phihat}) one can show that
\begin{align}
\hat{\phi}^j D_j \hat{\phi}_i = - D_i (\ln R) \, , 
\end{align}
and so
\begin{align}
\mathcal{R}^\Sigma = 2 \mathcal{K} - \frac{2}{R} \left( R_{TT} + R_{N\!N} \right)
\end{align}
with $R_{N\!N} = N^a\! N^b \!\bar{D}_a \bar{D}_b R$ and $R_{TT} = T^a \!T^b \!\bar{D}_a \bar{D}_b R$
following the notation of early sections. Applying  (\ref{HamCon}) this can be rewritten as
\begin{align}
\mathcal{K} = - \frac{1}{2} \mathcal{R} + \frac{R_{N\!N}}{R} 
+ \frac{1}{2} \left(k_{u}^2 - k_{u}^{ab}  k^u_{ab}  \right) + k_u k_{\phi \phi} \, , 
\end{align}
where 
\begin{align}
\mathcal{R}  &=   q^{ac} q^{bd} \mathcal{R}_{abcd} \label{R2} \\
&=   q^{ac}  T^b (d_a d_b T_c - d_b d_a T_c ) + 
q^{ac} \hat{\phi}^b \! (d_a d_b \hat{\phi}_c - d_b d_a \hat{\phi}_c ) \nonumber \\
& = - \frac{R_{TT}}{R} \nonumber
\end{align}
is the Ricci scalar for $S$ and we have used (\ref{Kex}). Note that this curvature is calculated for the 
connection associated with the two-dimensional metric $q_{AB}$. Hence $T^A d_A T^B = 0$ and
$\ddot{R} = T^A d_A (T^B d_B R) = R_{TT}$. 

Next, the two components of $N^a \! \bar{D}_a \kappa$ are
\begin{align}
N^a \! D_a \left(\frac{R_N}{R} \right)\eqg - \left( \frac{\dot{\beta}}{\beta} \right)\frac{\dot{R}}{ R} + \frac{R_{N\!N}}{R} - \frac{R_N^2}{R^2}
\end{align}
and
\begin{align}
N^a\!\bar{D}_a (K - K_{N\!N}) & \eqg - k^u_{ab} k_N^{ab} + k_N K_{N\!N} \, , \nonumber
\end{align}
where we have applied the diffeomorphism constraint (\ref{DiffCon}). 

Finally, substituting all of this into (\ref{ddb1}) we obtain
\begin{align}
0 \eqg &\,  \ddot{\beta} + \left( \! \frac{\dot{R}}{R} \right) \dot{\beta} 
 - \left(\frac{1}{2} \mathcal{R} - 2 | \sigma_+ |^2  \right) \beta   \, . \label{DeviationEq}
\end{align}
But in axisymmetry 
\begin{align}
\triangle_S \beta =  (T^A T^B \!+\! \hat{\phi}^A \hat{\phi}^B) d_A d_B \beta = \ddot{\beta} + \left( \! \frac{\dot{R}}{R} \right) \dot{\beta} 
\end{align}
and so the common solutions of (\ref{ddb1}) and (\ref{DeviationEq})  are exactly eigenvalue-zero eigenfunctions of the 
stability operator(\ref{eq:stability}).

Note: It is also possible 
(and quicker) to derive (\ref{DeviationEq}) directly from the constraint equations without going 
through (\ref{ddb1}) along the way. However the method that we present above demonstrates
the close relation to the deviation equations in a clearer way.

\subsection{Deviation equation as Sturm-Liouville problem}

We are primarily interested in closed \MOTSs and so MOTSodesics that intersect the $z$-axis. As we have seen, they
must do this at a right-angle. Then given a MOTSodesic $\gamma_o$ which intersects the $z$-axis at $s=0$ and $s=L$
we are interested in solutions to (\ref{DeviationEq}) on the interval $[0,L]$ with 
\begin{align}
\dot{\beta}(0)=\dot{\beta}(L) = 0 \; . \label{BC}
\end{align} 
Note that we have switched our curve parameter to $s$ since we are now focused on the central curve and 
also wish to avoid any confusion with the eigenvalues $\lambda$.

To better understand these ``nearby'' \MOTSs, rewrite the stability problem in Sturm-Liouville form:
\begin{align}
\dv{s} \left( R \, \dv{\beta}{s} \right) -
R  \left(\frac{1}{2} \mathcal{R} - 2 | \sigma_+ |^2  \right)   \beta = -\lambda \beta  \; . \label{DeviationEqSL}
\end{align}
Since $R(0)=R(L)=0$, this is a singular Sturm-Liouville problem. However  we expect that its eigenfunctions $\beta_n$ satisfy the standard properties 
for eigenfunctions of regular Sturm Liouville operators\footnote{Despite some searching and consultations we haven't been able to identify a 
published theorem that guarantees this. However (\ref{DeviationEqSL}) has a similar form to many of the singular problems of mathematical physics that do have these properties. As in those 
 problems, the singularity results from the polar-type coordinates. Further we observe all of these properties in the upcoming sections. Hence for now 
 we assume that they are true.}.
That is:
\begin{enumerate}[i)]
\item Their eigenvalues $\lambda_n$ are all real and non-degenerate.
\item There is a smallest eigenvalue $\lambda_0$ but no largest one. 
\item The $\beta_n$ form an orthogonal basis on $\mathcal{L}^2 (0,L)$.
\item Ordering the $\beta_n$ by the value of their eigenvalues (and starting the labelling at $n=0$), 
$\beta_n$ has exactly $n$ zeros in $(0,L)$. 
\end{enumerate}
The eigenvalues here are those of (\ref{DeviationEqSL}) rather than the full stability operator. Relative
to the full operator $\lambda_n = \lambda_{n,0}$. 

Then given a MOTS, the stability characterization determines the properties of solutions to the deviation equation 
that satisfy (\ref{BC}). If a MOTS $S_o$ is 
\begin{enumerate}[a)]
\item \emph{strictly stable} ($\lambda_0>0$): there is no such $\beta(s)$. 
\item \emph{stable but not strictly stable} ($\lambda_0 = 0$): there is a single $\beta(s)$. 
It has no zeros on $(0,L)$. 
\item \emph{unstable with vanishing eigenvalue} ($\lambda_n = 0$ for some $n\neq 0$): there is a single $\beta(s)$.
It has $n$ zeros on $(0,L)$. 
\item \emph{unstable, no vanishing eigenvalue} ($\lambda_n \neq 0$ for any $n$): there is no such $\beta(s)$. \end{enumerate}

Intuitively we then have the following picture (in axisymmetry). 
MOT(O)S close to a stable MOTS cannot intersect it. However if it is unstable we expect the nearby MOT(O)S to intersect
it a number of times equal to the number of negative eigenvalues of the (\ref{DeviationEqSL}). This assigns a direct
geometrical meaning to the number of negative eigenvalues of the stability operator. 
Keep in mind however that these conclusions are 
all in the linearized regime of the deviation equation. For ``full'' MOTSodesics we will expect some modifications to these
behaviours. In particular we probably don't expect continuous families of closed MOTSodesics, even for those curves 
with vanishing eigenvalues. 

%
%
%
%
%

In retrospect the identification of zero-eigenvalue eigenfunctions of the stability operator with solutions to the MOTSodesic
equation should not have been surprising. For geodesics, zero-eigenvalue eigenfunctions of the Jacobi operator
are solutions to the geodesic deviation equation and the number of conjugate points along the geodesic depends 
on the number of negative eigenvalues according to the Morse Index Theorem \cite{spivak4}. Strictly stable \MOTSs are the 
analogue of geodesics in a negatively curved (hyperbolic) background, stable \MOTSs are the analogue 
of geodesics in a vanishing curvature (flat) background and unstable ones are the analogue of geodesics in a 
positively curved background. We will return to  this
in more detail in a future work but for now consider examples.

%
%
%
%

%
%

%



\subsection{Example: Flat spacetime}
The simplest example to consider is a $t=\mbox{constant}$ surface in Minkowski spacetime.  Then
\begin{align}
h_{ij} \dd x^i \dd x^j = \dd \rho^2 + \dd z^2 + \rho^2 \dd \phi^2
\end{align}
and the extrinsic curvature $K_{ij} = 0$. Then there is no distinction between outward and inward 
oriented \MOTSs and, in fact, they are axisymmetric minimal surfaces in Euclidean $\mathbb{R}^3$: that 
is catenoids and $z=\mbox{constant}$ planes. We first confirm that these are solutions of the 
MOTSodesic equations. 

\subsubsection{MOTSodesics}

For these surfaces $R = \rho$ and the metric in $\bar{\Sigma}$ is
\begin{align}
\bar{h}_{ab} \dd x^a \dd x^b = \dd \rho^2 + \dd z^2  \label{h2} \; . 
\end{align}
Hence the Christoffel symbols vanish. We further have $R=\rho$ and 
\begin{align}
N = - \dot{Z} \dv{\rho} + \dot{P} \dv{z} \; . \label{Nvect}
\end{align}
Then the MOTSodesic equations \ref{MainEq} become
\begin{align}
\ddot{P} & =\phantom{-} \frac{\dot{Z}^2}{P} \label{FlatEq}\\
\ddot{Z} & = - \frac{\dot{P}\dot{Z}}{P} \, ,  \nonumber
\end{align}
since $k_u = 0$ and so $\kappa = R_N = - \frac{\dot{Z}}{P}$. This $\kappa$ works as a repulsion
term that tries to divert these curves from hitting the $z$-axis. If $\dot{Z} \neq 0$ during the
approach then $\kappa \rightarrow \infty$ as $\rho \rightarrow 0$. 

These equations have a trivial solution:
\begin{align}
P_{\mbox{\tiny{plane}}} &= \rho_o - s \label{plane}\\
Z_{\mbox{\tiny{plane}}} &= z_o  \, , \nonumber
\end{align}
where $\rho_o = P(0)$ and we have chosen to orient $s$ so that positive moves towards the 
axis (reaching it at $s=\rho_o$). 
These are the planes of constant $z$. 

The catenoid solutions are not so obvious. 
With respect to arclength a catenoid may be parameterized as 
\begin{align}
P_{\mbox{\tiny{cat}}} &= \sqrt{\rho_o^2 + (s-s_o)^2}  \label{catenoidSol} \\
Z_{\mbox{\tiny{cat}}} &= z_o + \rho_o\,\mbox{arcsinh} \left( \frac{s-s_o}{\rho_o} \right) \nonumber
\end{align}
where $(\rho_o, z_o)$ is the closest approach to the $z$-axis, which happens at $s=s_o$. 
It is straightforward to confirm that these are indeed
solutions to (\ref{FlatEq}). 

\begin{figure}
\includegraphics{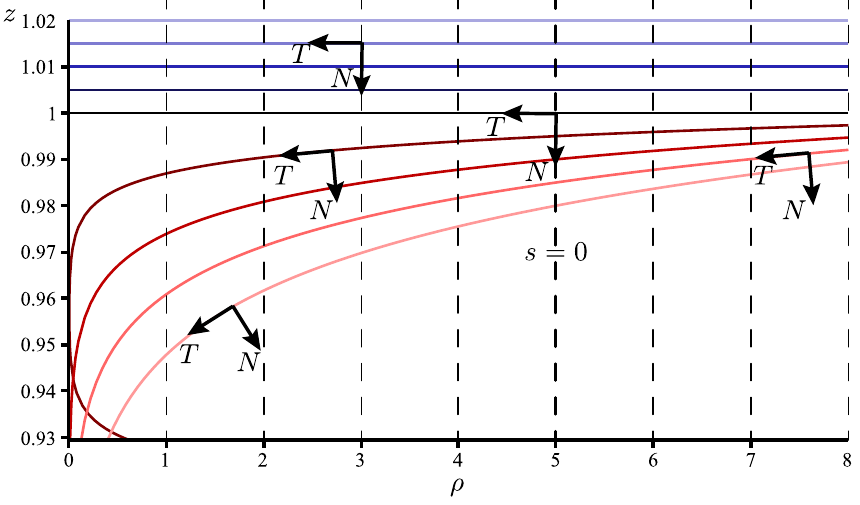}
\caption{A family of MOTSodesics in flat space. For $z \geq 1$ they rotate into planes while below
that line they rotate into catenoids. Note that in the bottom left-hand corner MOTSodesics intersect
and so the congruence-associated coordinate system fails. For small $\beta$, its value is approximately
the distance measured from $z=1$ to the nearby curves along the dashed lines of constant $s$.   }
\label{FlatMOTS}
\end{figure}

A family of MOTSodesics is shown in FIG.~\ref{FlatMOTS} with planes above $z=1$ and catenoids 
below. Within the range of the figure, the only catenoid that can be seen to have fully turned around
is the dark red one that started closest to $z=1$. 

\subsubsection{MOTSodesic deviation}

Next consider the MOTSodesic deviation equation, though here of course we are considering \MOTOSs rather than \MOTSs
since none of these surfaces close. For these non-compact cases many of the Sturm-Liouville conclusions will not 
apply but this case can still demonstrate some properties of the equation. In particular we can use it to better 
understand the repulsion from the $z$-axis.

From (\ref{R2}) and (\ref{FlatEq})  the deviation equation (\ref{DeviationEq}) becomes
\begin{align}
\ddot{\beta} + \left( \! \frac{\dot{P}}{P} \! \right) \dot{\beta} - 2 \left(\! \frac{\dot{Z}}{P} \! \right)^2 \beta = 0 \; .
\label{FlatDev} 
\end{align}
This has exact solutions for both planes and catenoids but
for purposes of this example, it will be sufficient to 
consider the solution for planes. Then $\dot{Z} = 0$ and so $\beta = \beta_o$ is an eigenvalue-zero eigenfunction. 
Hence it is stable but not strictly stable.

More generally,
\begin{align}
\beta_{\mbox{\tiny{plane}}} =  \beta_o - \dot{\beta}_o \rho_o \ln \left( 1 - \frac{s}{\rho_o} \right) 
\end{align}
where $\beta_o = \beta(0)$ and $\dot{\beta}_o = \dot{\beta}(0)$. In this case (and only this case) 
we have chosen $s=0$ to be a point not on the $z$-axis as we want to use this as a model to demonstrate how curves
diverge during that approach. 

If $\dot{\beta}_o = 0$ then the curves remain parallel. These are in the parallel line
above $z=1$ in FIG.~\ref{FlatMOTS}. These rotate into the planes in $\Sigma$ and are the
only curves that never intersect $z=1$ (as one would expect from Euclidean geometry!). 

If $\dot{\beta} \neq 0$ then $\beta$ monotonically increases in the approach to $z=0$ until it diverges as 
$s \rightarrow \rho_o$. This effect can be seen for the $z<1$ catenoids.  Meanwhile as $s \rightarrow -\infty$
$\beta$ monotonically decreases and all of these curves ultimately cross $z=1$. 

These exact solution behaviours are consistent with the expected properties of solutions of
 (\ref{DeviationEq}). In particular, this simplest example demonstrates how only 
 finely tuned congruences of MOTSodesics can intersect the $z$-axis. 
Generic curves turn back and in particular, even if one element reaches the axis, its 
neighbours will usually dramatically diverge from it.

%

%
%

\subsection{Example: Schwarzschild-Painlev\'e Gullstrand}
\begin{figure*}
\includegraphics{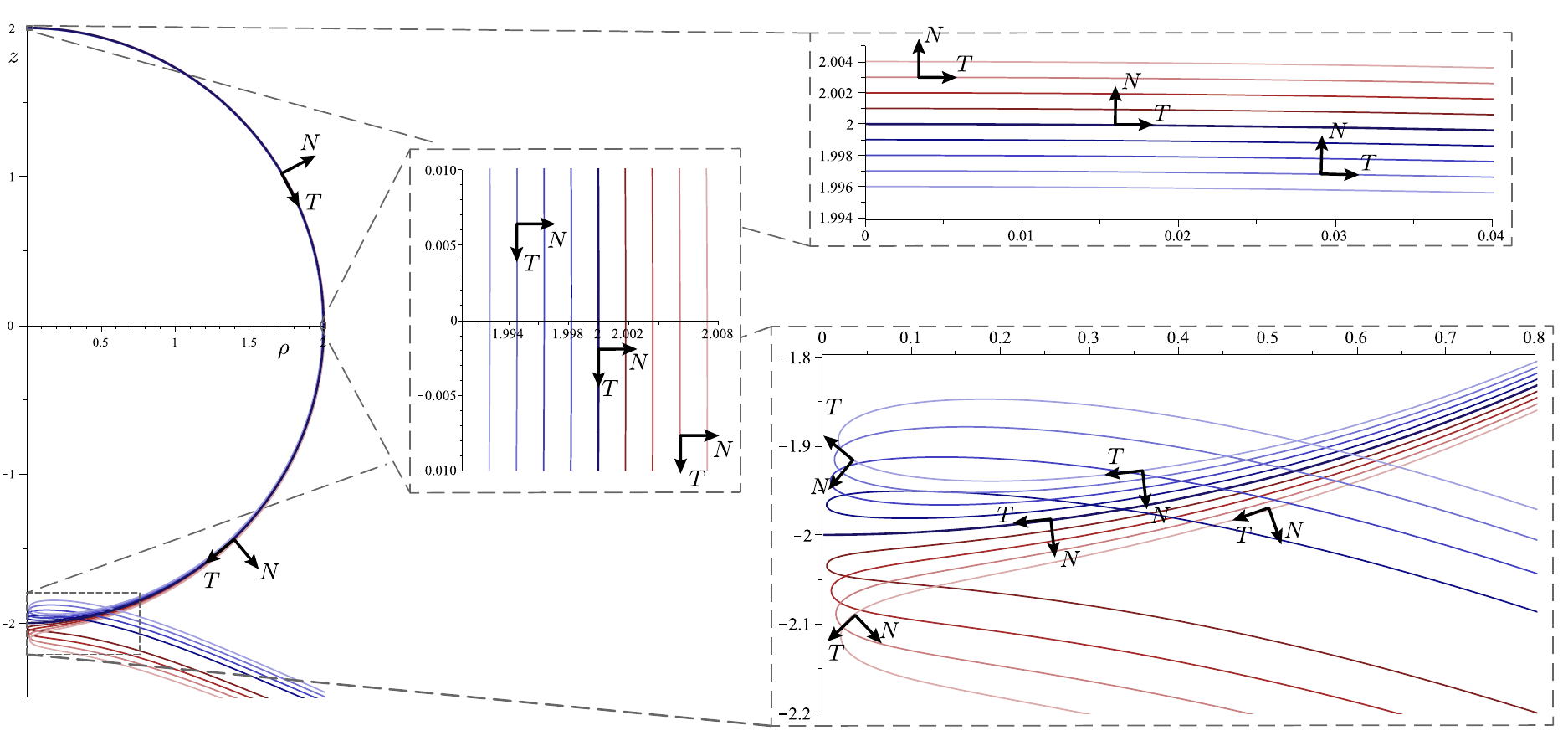}
\caption{MOTSodesics near the Schwarzschild apparent horizon in a slice of constant Painelev\'e Gullstrand time. Note the relatively slow divergence until the curves approach the $z$-axis. The axes have units of mass $M$.  }
\label{MOTSodesicEx}
\end{figure*}

The plane $z=1$ considered in the last section was stable ($\lambda_0 = 0$) but not strictly stable. Further
it was a MOTOS rather than MOTS. We now consider a strictly stable MOTS: the Schwarzschild horizon. 

We work in Painlev\'e-Gullstrand coordinates, the case for which the \MOTSs were studied in detail in 
\cite{Booth:2020qhb}. Then the induced metric is still flat
\begin{align}
h_{ij} \dd x^i \dd x^j = \dd \rho^2 + \dd z^2 + \rho^2 \dd \phi^2 \label{Flat3D}
\end{align}
while the extrinsic curvature is
\begin{align}
K_{ij} \dd x^i \dd x^j =  \sqrt{\frac{M}{2}} \Bigg(& \frac{\rho^2 - 2 z^2}{ r^{7/2}} \dd \rho^2 + 
\frac{6 \rho z}{r^{7/2}} \dd \rho \, \dd z \label{KSchw} \\
& + \frac{z^2-2 \rho^2}{ r^{7/2}} \dd z^2 - \frac{2\rho^2}{r^{3/2}} \dd \phi^2 \Bigg) \nonumber
\end{align}
where $r=\sqrt{\rho^2 + z^2}$ is the regular radial spherical coordinate. 

\subsubsection{MOTSodesics}
For these surfaces we still have $R=\rho$, a flat metric (\ref{h2}) on $\bar{\Sigma}$ and normal vector
(\ref{Nvect}). However 
\begin{align}
k_u = -\sqrt{\frac{M}{2}} \left(    \frac{3 (Z \dot P - P \dot Z)^2}{r^{7/2}} + \frac{1}{r^{3/2}} \right)
\end{align}
does not vanish. 

We again have $R =\rho$ and so $R_N = -\frac{\dot{Z}}{P}$. However 
\begin{align}
\kappa =  -\frac{\dot{Z}}{P} \pm k_u
\end{align}
and so differs for a left versus right-oriented MOT(O)S. Nevertheless the MOTSodesic equations
still take a relatively simple form:
\begin{align}
\ddot{P} & =\phantom{-} \frac{\dot{Z}^2}{P} \pm k_u \dot{Z} \label{PGEq}\\
\ddot{Z} & = - \frac{\dot{P}\dot{Z}}{P} \mp k_u \dot{P} \; .  \nonumber
\end{align}
The MOTS at $r=2M$ is parameterized as
\begin{align}
P & = 2 M \sin \left( \frac{\lambda}{2M} \right) \\
Z & = 2 M \cos \left( \frac{\lambda}{2M} \right) \nonumber
\end{align}
and can easily be checked to be a solution, but in general these equations cannot be solved
exactly. However they are very easily solved numerically using standard ODE solvers and some consistently oriented
near horizon MOTSodesics are shown in FIG.~\ref{MOTSodesicEx}. More exotic ones are shown in  FIG.~\ref{PGExotic}.

\subsubsection{MOTSodesic deviation}
The deviation equation becomes
\begin{align}
0 \eqg &\,  \ddot{\beta} + \left( \! \frac{\dot{P}}{P} \right) \dot{\beta} 
 + \left( \frac{\ddot{P}}{P}  + 2 | \sigma_+ |^2  \right) \beta   \;  . 
\end{align}
%
In general this doesn't have an exact solution but there is an exception for the deviation of MOTSodesics from 
$r=2M$. Then the deviation equation becomes:
\begin{align}
\ddot{\beta} +\cot \left( \frac{\lambda}{2M} \right)  \frac{\dot{\beta}}{2M}  - \frac{\beta}{4M^2}  = 0 \label{SchwDev}
\end{align}
which has general solution
\begin{align}
\beta = & A P_{\frac{-1+i\sqrt{3}}{2}} \left(\cos \left( \frac{\lambda}{2M} \right) \right)  \label{SchwarzDevSol}\\
& + B P_{\frac{-1+i\sqrt{3}}{2}} \left(\cos \left(\pi- \frac{\lambda}{2M} \right) \right) \nonumber
\end{align}
where $P_{\frac{-1+i\sqrt{3}}{2}}$ is a Legendre function and $A$ and $B$ are arbitrary constants. 
We are interested in these solutions in the range $0 \leq \lambda \leq 2M \pi$. The term first is finite
and has derivative $0$ at $\lambda =0$ but diverges at $\lambda = 2M \pi$. The second term has
the opposite behaviour.  
For the case shown in FIG.~\ref{MOTSodesicEx}, $A = \beta_o$ and $B=0$ while these would be
reversed for the equivalent set of MOTSodesics launched from the south pole. For cases where the 
MOTSodesics are instead parallel at the equator (or any other non-polar point), $\beta$ diverges at
both ends.

Note that while these are solutions of (\ref{SchwDev}) they are not solutions of the full Sturm-Liouville 
problem: they always diverge at either one or both ends. However this is not a surprise: $r=2M$ is a strictly
stable MOTS and so one does not expect the full problem to have any solutions. Instead, 
the separation between initially parallel MOTSodesics monotonically increases just as initially parallel geodesics
diverge in a hyperbolic background. The rate of growth for that separation  
quantitatively matches that of (\ref{SchwarzDevSol}) with $A=2M$ and $B=0$ (until the final stages of the divergence). 

\begin{figure}
\includegraphics{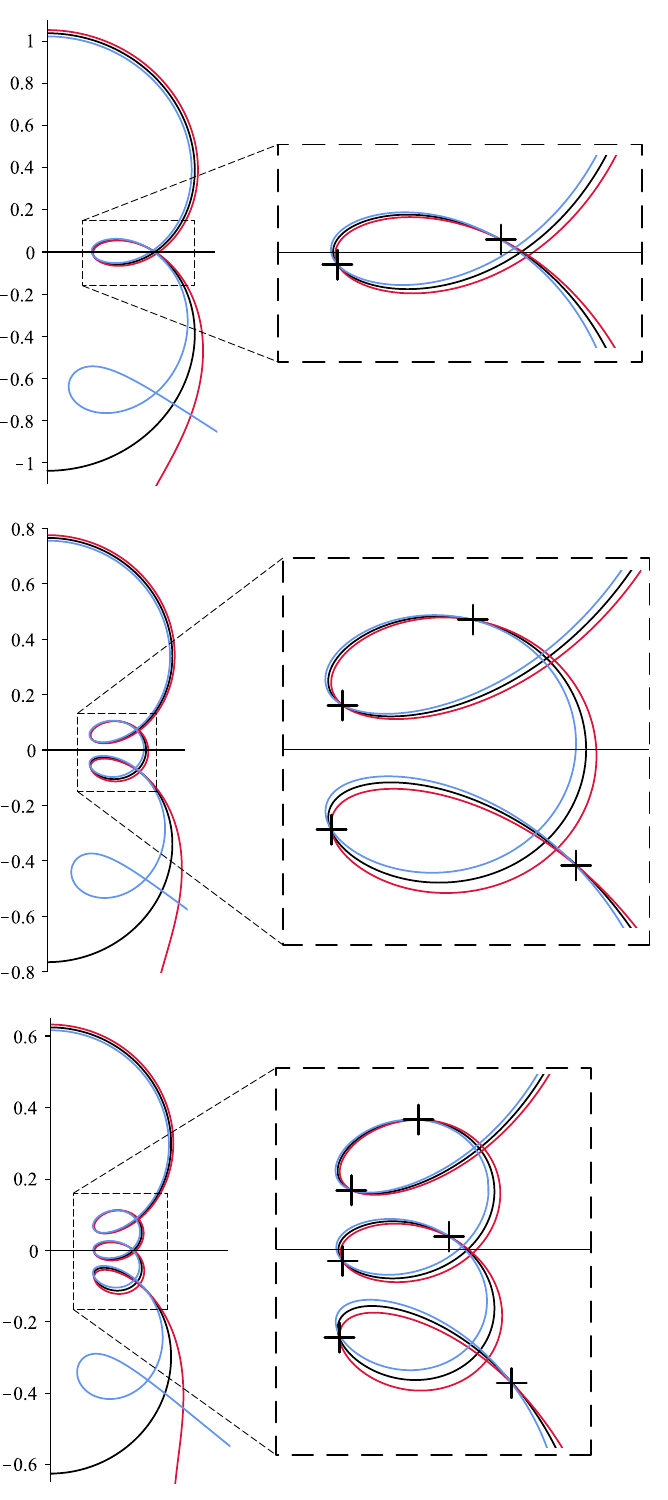}
\caption{The first three self-intersecting Schwarzschild MOTSodesics. They are unstable with the number of negative
eigenvalues equal to twice the number of loops. Conjugate points are marked with $+$-signs. }
\label{PGExotic}
\end{figure}

In contrast to this strictly stable MOTS, the self-intersecting \MOTSs from \cite{Booth:2020qhb} are unstable and 
from numerical experiments we find that the stability operator has two negative eigenvalues per loop of the MOTSodesic. 
This behaviour is show in FIG.~\ref{PGExotic}. Note that initially parallel MOTSodesics oscillate around the central
curve before ultimately diverging in the approach to the $z$-axis (where they leave the linear regime and so the
deviation equation no longer applies).

%

%
%
%
%
%
%
%
%

\subsection{Example: Reissner-Nordstr\"om in PG coordinates}

A simpler example of unstable MOTSodesics appeared in \cite{Booth:2017fob}. 
Consider the inner horizons of Reissner-Nordstr\"om spacetimes. 
In Painlev\'e-Gullstrand coordinates for these solutions, the induced metric on $\Sigma$ remains flat (\ref{Flat3D}) while the 
extrinsic curvature becomes
\begin{align}
K_{ij} \dd x^i \dd x^j =  \sqrt{\frac{M}{2} \!-\! \frac{Q^2}{4r}} \Bigg(& \! \frac{\rho^2 - 2 z^2}{ r^{7/2}} \dd \rho^2 + 
\frac{6 \rho z}{r^{7/2}} \dd \rho \, \dd z \nonumber  \\
& + \frac{z^2-2 \rho^2}{ r^{7/2}} \dd z^2 - \frac{2\rho^2}{r^{3/2}} \dd \phi^2 \! \Bigg) \; . 
\end{align}
These coordinates cover both the outer and inner horizons though they fail for $r < \frac{q^2}{2m}$ (which is always
inside the inner horizon). 

The MOTSodesic equation for these solutions is unchanged from (\ref{PGEq}) modulo the extrinsic curvature term 
which becomes
\begin{align}
k_u = -\sqrt{  \frac{M}{2}-\frac{Q^2}{4r} } \left(    \frac{3 (Z \dot P - P \dot Z)^2}{r^{7/2}} + \frac{1}{r^{3/2}} \right) \; . 
\end{align}
Again the general solution cannot be written down in closed form, however one can check that the 
outer horizon $r_{\mbox{\tiny{out}}} = M + \sqrt{M^2 - Q^2}$ and inner horizon $r_{\mbox{\tiny{in}}} = M - \sqrt{M^2 - Q^2}$ 
are solutions. 

These solutions are non-vacuum and so the stability operator includes matter terms. From \cite{Booth:2017fob} on either 
the outer or inner horizon it takes the form:
\begin{align}
L_\Sigma \psi = -\dd^2 \psi +(r F') \psi  \; ,
\end{align}
where $F = 1 - \frac{2M}{r} + \frac{Q^2}{r^2}$. The eigenfunctions are the spherical harmonics $Y_{lm} (\theta, \phi)$ and so for the inner horizon, $L_\Sigma$ has (degenerate) eigenvalues 
\begin{align}
\lambda_{l,m} = l(l+1) - \frac{2\sqrt{M^2-Q^2}}{M-\sqrt{M^2-Q^2}} \, , 
\end{align} 
for $l \in \mathbb{Z}^+$. 
The principal eigenvalue is then 
\begin{align}
\lambda_{0,0} = - \frac{2\sqrt{M^2-Q^2}}{M-\sqrt{M^2-Q^2}} \; . 
\end{align}
This vanishes for an extremal horizon $M=Q$ (for which  $r_{\mbox{\tiny in}} = r_{\mbox{\tiny out}}$) but is otherwise 
negative. Hence for $Q<M$ the inner horizon is unstable. 

\begin{figure}
\includegraphics[scale=1.2]{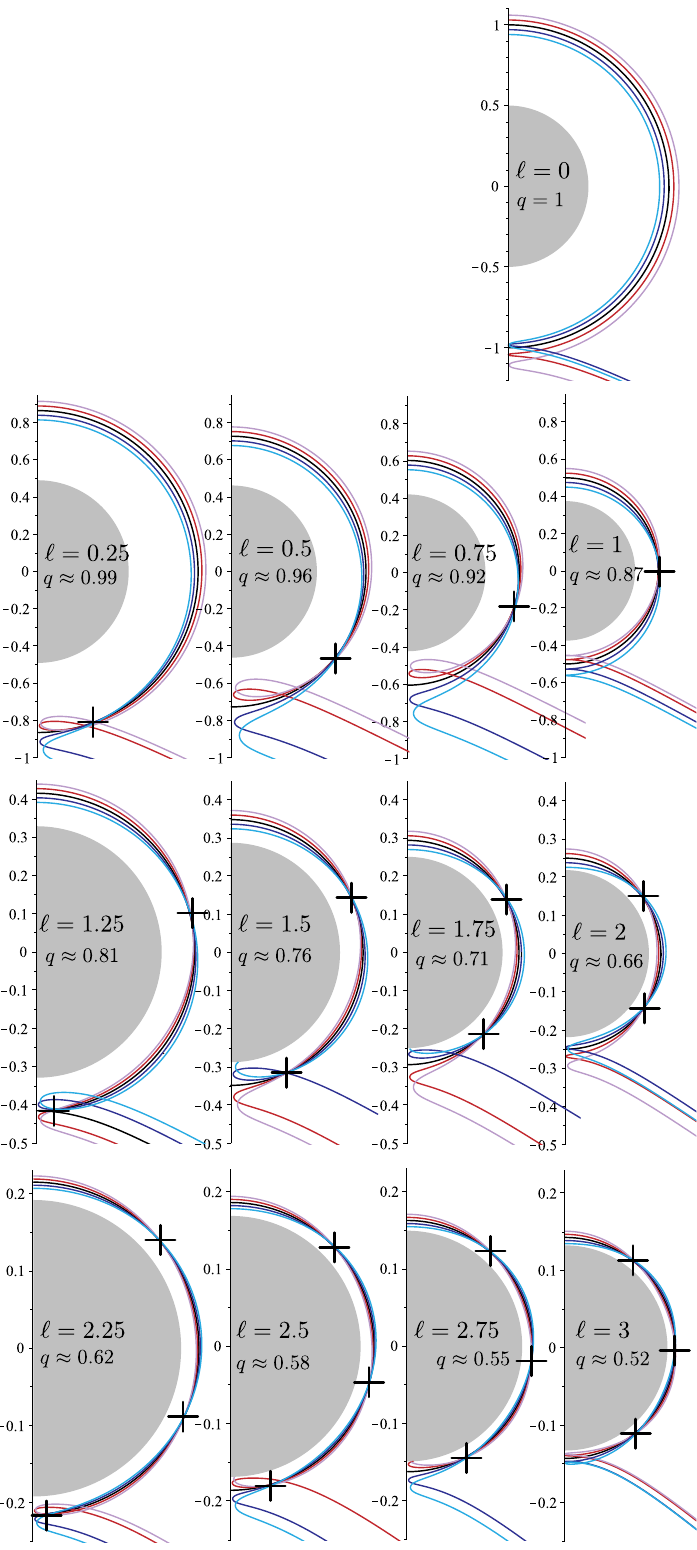}
\caption{Inner spherically symmetric MOTS from the Reissner-Nordstr\"om spacetime. $Q$ is chosen from (\ref{Qx}) setting $l_o = \ell$. 
Thus the $m=0$ version of the stability operator has zero negative eigenvalues in the first row, one in the second, two in the third and three
in the fourth. 
The correspondence between the stability operator and deviation equation can then be clearly seen in the number
of intersections (again marked with $+$s) between the nearby MOTSodesics and the inner horizon. 
The gray circle indicates the region  not covered by the Painlev\'e-Gullstrand coordinates. Note that some MOTSodesics
are lost into that region.  }
\label{RNEx}
\end{figure}

Any of the other eigenvalues may also be made to vanish by a careful choice of the physical parameters. Specifically 
for a given $l_o$ if we choose
\begin{align}
Q = \pm \frac{2M \sqrt{l_o^2+l_o+1}}{l_o^2+l_o+2}  \label{Qx}
\end{align}
we have 
\begin{align}
r_{\mbox{\tiny in}} = \frac{2M}{l_o^2 + l_o + 2}
\end{align}
and the eigenvalues $\lambda_{l_o,m} = 0$. More generally choosing
\begin{align}
 \frac{2M \sqrt{l_o^2+3l_o + 3}}{l_o^2 + 3l_o + 4} < |Q| \leq \frac{2M \sqrt{l_o^2+l_o + 1}}{l_o^2 + l_o + 2} \label{qbound}
\end{align}
the stability operator will have $l_o$ axisymmetric eigenfunctions $Y_{l0}$ with negative eigenvalues\footnote{Of course
each of these eigenvalues is actually $(2l+1)$-times degenerate with non-axisymmetric functions with  $-l \leq m \leq l$ for $0 \leq l\leq l_o$.}.

We directly derived the deviation operator only in vacuum, however it is straightforward to show that the stability 
operator/MOTSodesic deviation correspondence continues to hold in the presence of matter. Then from the deviation equation we would 
expect that for $Q$ bound as in (\ref{qbound}), initially parallel MOTSodesics that are sufficiently close to the inner horizon
would cross it $l_o$ times. This is the case, as demonstrated by FIG.~\ref{RNEx}. 

Given an initial unstable MOTSodesic, sufficiently nearby initially parallel MOTSodesics will oscillate around it. 
However this linearized behaviour is overcome by the intense repulsion from the $z$-axis when the congruence 
approaches the south pole. Hence in the figure the ``inside'' curves end up with an extra intersection that happens
after the repulsion (and so is outside the linearized regime of the deviation equation). 

Note too that while the Sturm-Liouville only told us about numbers of intersections for eigenfunctions we can see
from this example that even for the unstable inner horizons with no vanishing eigenvalues, the number of intersections
corresponds to the number of negative (degenerate) eigenvalues. This is in line with Morse index theory for geodesics,
where the number of negative eigenvalues of the Jacobi operator corresponds to the number of conjugate points 
along a curve\cite{spivak4}, however we have not yet rigorously extended those results to MOTSodesics. 

\subsection{Example: Brill-Lindquist Initial Data}
\label{sec:return_to_BL}
Finally we return to the Brill-Lindquist initial data. Here neither the geodesic nor the deviation equation can be solved
exactly and so we consider numerical solutions. In particular the eigenvalue spectrum and  so the number of negative 
eigenvalues is calculated using the methods of \cite{pook-kolb2020I,pook-kolb2020II}. 

\begin{figure*}
\includegraphics[width=0.3\textwidth]{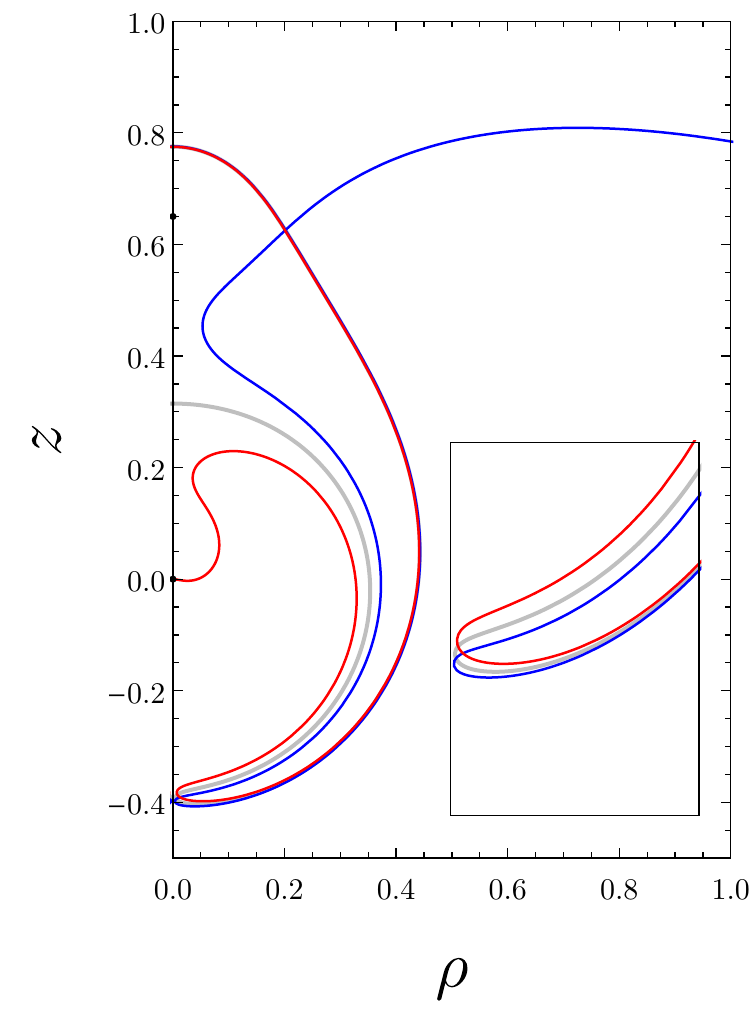}
\quad
\includegraphics[width=0.3\textwidth]{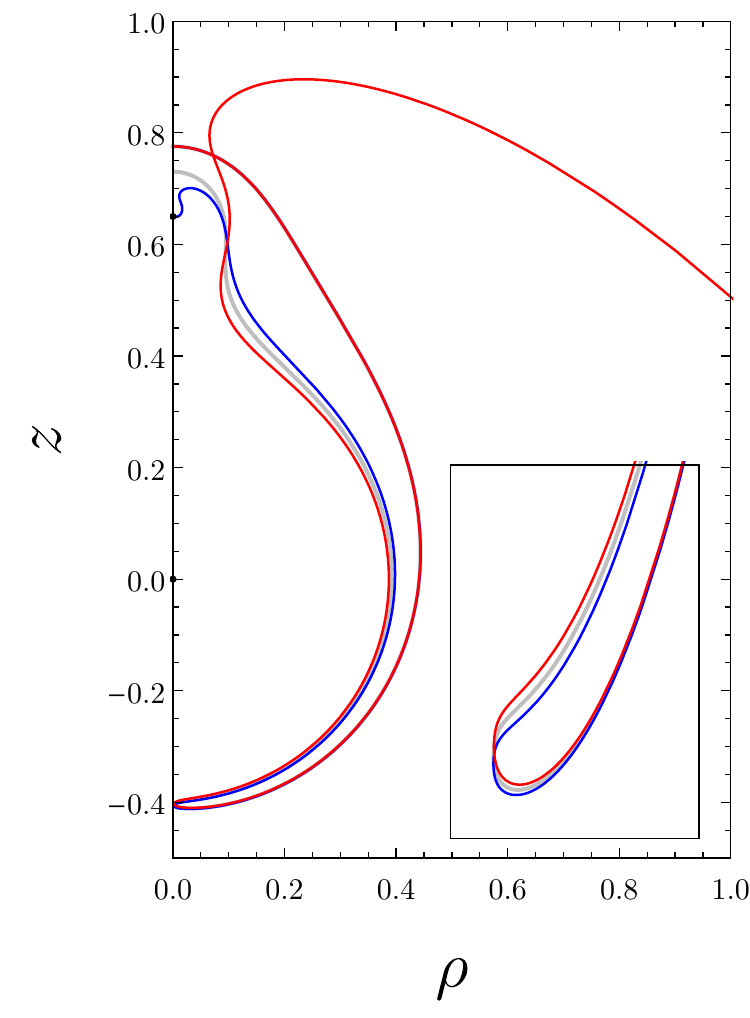}
\quad
\includegraphics[width=0.3\textwidth]{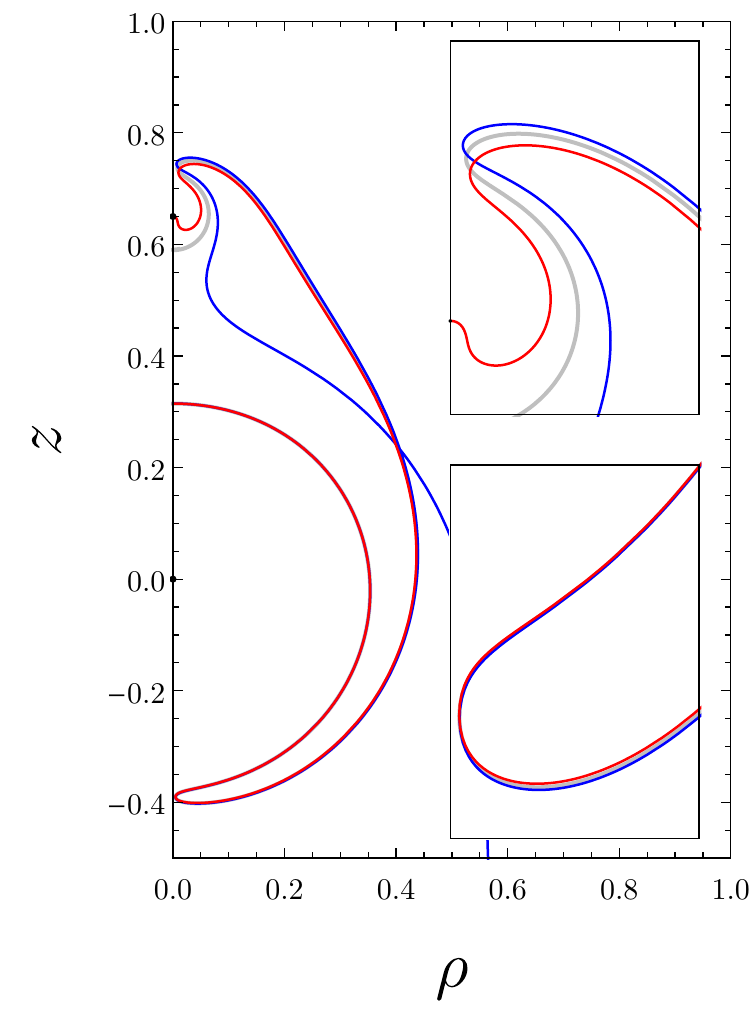}
\caption{Nearby MOTSodesics for three of the new MOTSs presented in Figure~\ref{fig:moreMOTS}. In each case, the gray curve corresponds to the MOTS, while the red curve originates slightly inside the MOTS and the blue curve slightly outside. For the MOTS shown here, we have determined that the stability operator has 1, 2 and 2 negative eigenvalues, respectively. We can see that this corresponds also to the number of intersections of nearby MOTSodesics. The insets show a zoomed-in view near the intersections.
\label{BLDev}}
\end{figure*}

Nevertheless as shown in FIG.~\ref{BLDev} the results observed for exact solutions continue to hold. 
The number of negative eigenvalues of the 
stability operator corresponds to the number of intersections between initially parallel curves (in the regime while
the linear approximation can be assumed to hold). 

From these examples we can also understand the geometric meaning of these negative eigenvalues. A turn from the 
$z$-axis generates a negative eigenvalue. This can be understood as resulting from the fact that, to leading order, 
the shape of the turn of nearby MOTSodesics from the $z$-axis is necessarily the same. Then at each such turn the 
inside and outside MOTSodesics will necessarily switch places. This effect can be seen in all three subfigures. 

A negative eigenvalue also appears to arise when a MOTSodesic switches from ``orbiting'' from one stable MOTS to 
another. Hence the green inner MOTS in FIG.~\ref{fig:BL-standard_horizons} picks up one negative eigenvalue as does 
the middle subfigure of FIG.~\ref{BLDev}. 

These two counting rules appear to be sufficient to account for all negative eigenvalues seen in Brill-Lindquist 
initial data MOTS and also appear to hold for the more general MOTS in the sequel \cite{PaperII}. 
However at this time we do not have a rigorous demonstration of this.

\section{Conclusions}
\label{sec:conclusions}

In the present paper we have developed a new generalized shooting method for
finding marginally outer trapped (open) surfaces in axisymmetric and
non-spinning but otherwise
arbitrary initial data.
This has led to the discovery of a large number of previously unknown \MOTSs 
in even the simplest spacetimes. Given this explosion of examples we have chosen
to restrict the term ``horizon'' to stable \MOTSs. Unstable \MOTSs (and their associated MOTTs)
should not be interpreted as black hole boundaries. 

It is certainly reasonable to wonder whether these new types of \MOTSs
continue to exist during a dynamical evolution of initial data or whether these are only features of 
(time-symmetric)
initial data. The second paper in this series\cite{PaperII} unequivocally demonstrates that not only do
similar \MOTSs exist during mergers but also they play a key role in understanding the 
final fate of the initial apparent horizons. \MOTSs dynamically form and annihilate in pairs
and ultimately the initial apparent horizons are annihilated by more exotic \MOTSs. 

Faced with this plethora of \MOTSs we developed the MOTSodesic deviation equation to better understand
how they relate to each other. That investigation revealed the retrospectively obvious result that the stability
operator for \MOTSs is the analogue of the stability operator for geodesics.
This has then provided a new insight into  
stable and unstable \MOTSs and the geometric implications of negative eigenvalues of the stability operator. 
We expect that these results can all be rigorously proved using methods very similar to those used for the analogous
proofs for geodesics and the Jacobi operator. However that is beyond the scope of the current paper. 

It will  be fascinating to see if and how these observations extend beyond axisymmetry. 
In that general case, eigenvalues
of the stability operator (beyond the principal eigenvalue) are generally complex and one cannot expect to simply count
negative eigenvalues. Further zeros of the eigenfunctions would be expected to occur along curves rather than at
points and so we expect any analogous results would need to be phrased in terms of nodal domains. However 
even for minimal surfaces such results are much less straightforward than for geodesics and so we do not expect
the general case to be resolved quickly! That said, as in this series of papers one may hope that a combined 
theoretical and numerical investigation may obtain unexpected results and insights.

\begin{acknowledgments}
    We would like to express our gratitude to 
    Graham Cox, 
    Jose~Luis~Jaramillo,
    Badri~Krishnan, Hari Kunduri and the members of the 
    Memorial University Gravity Journal Club
    for valuable discussions and suggestions.
 
   IB was supported by the Natural Science and Engineering Research Council of Canada Discovery Grant 2018-0473. 
   The work of RAH was supported by the Natural Science and Engineering Research Council 
   of Canada through the Banting Postdoctoral Fellowship program and also by AOARD Grant FA2386-19-1-4077.

\end{acknowledgments}

\appendix

\section{$\kappa$ on  the  $z$-axis}
\label{AppKappa}

Here we calculate the curvature $\kappa$ and  along a curve that intersects the $z$-axis. This is a tool in finding the series expansion
of these curves. To save space and make equations more readable we use the following notation:
$R_T = T^a \bar{D}_a R$, $R_N = N^a \bar{D}_a R$, $R_{TN} = T^a N^b \bar{D}_a \bar{D}_b R$.

The key equations used in these derivations are:
\begin{align}
T^b \bar{D}_b T^a = \phantom{-} \kappa N^a \\
T^b \bar{D}_b N^a = - \kappa T^a \nonumber \; . 
\end{align}
So, with overdots denoting derivatives with respect to $s$,
\begin{align}
\dot{R}_T & = \kappa R_N + R_{TT} \\
\dot{R}_N & = - \kappa R_T + R_{TN} \; .
\end{align}

Then if $s=0$ along the $z$-axis ($\rho = 0$) we are interested in the limit:
\begin{align}
 \kappa^\pm_o \equiv  \lim_{s \rightarrow 0}  \left( \frac{R_N}{R}  \pm  k_u \right)
\end{align}
with the first term being of the form $\frac{0}{0}$ as $s \rightarrow 0$. We can apply l'H\^{o}pital's rule:
\begin{align}
\kappa_o^\pm & = \lim_{s \rightarrow  0} \left(  \frac{R_N}{R}  \pm  k_u \right) \\
& = \lim_{s \rightarrow 0} \left(  \frac{\dot{R}_N}{\dot{R}}  \pm  k_u \right) \nonumber \\
& =  \lim_{s \rightarrow 0} \left(  \frac{- \kappa^\pm R_T + R_{TN} }{R_T}  \pm  k_u \right) \nonumber
\end{align}
and so rearranging and solving for $\kappa_o^\pm$ we find
\begin{align}
\kappa_o^\pm \eqz \frac{1}{2} \left(\frac{R_{TN}}{R_T} \pm k_u \right) \label{kappa0}
\end{align}
where as in the main text, the overset $z$ indicates that the righthand side should be evaluated on the $z$-axis.

\bibliography{blmotos}{}

\end{document}